\documentclass[
  aps,
  prb,
  10pt,
  twocolumn,
  superscriptaddress,
  longbibliography,
  nofootinbib
]{revtex4-2}

\usepackage[T1]{fontenc}
\usepackage[utf8]{inputenc}
\usepackage{amsmath,amssymb,amstext,enumerate,graphicx,physics,extarrows,mathtools}
\usepackage[nointegrals]{wasysym}
\SetSymbolFont{wasy}{bold}{U}{wasy}{m}{n}
\usepackage{bm}
\usepackage{graphicx}
\usepackage{booktabs}
\usepackage{xcolor}
\usepackage{silence}
\usepackage{etoolbox}
\usepackage[colorlinks=true,linkcolor=blue,citecolor=blue,urlcolor=blue]{hyperref}
\usepackage{orcidlink}

\makeatletter
\patchcmd{\@bibdataout@aps}{author="08"}{author="48"}{}{}
\patchcmd{\@bibdataout@aps}{author="08"}{author="48"}{}{}
\makeatother

\newcommand{\kvec}{\vb*{k}}
\newcommand{\rvec}{\vb*{r}}
\newcommand{\ii}{\mathrm{i}}
\newcommand{\ee}{\mathrm{e}}

\makeatletter

\newcommand{\Rmnum}[1]{\expandafter\@slowromancap\romannumeral #1@}
\makeatother

\def\nearestbracket#1{\mathinner{\left\langle{#1}\right\rangle}}
\def\nnb{\nearestbracket}

\graphicspath{{figures/}{img/}}

\begin{document}

\title{Hidden variable \texorpdfstring{$\mathbb{Z}_2$}{Z2} topological order
in a bilayer kagom\'e antiferromagnet}

\author{Hao Tian$^{\orcidlink{0009-0007-2238-0883}}$}
\affiliation{International Center for Quantum Materials, Peking University, Beijing 100871, China}

\author{Pengwei Zhao}
\affiliation{Department of Physics, Hong Kong University of Science and Technology, Clear Water Bay,
Hong Kong, China}
\affiliation{International Center for Quantum Materials, Peking University, Beijing 100871, China}

\author{Gang v.~Chen}
\email{chenxray@pku.edu.cn}
\affiliation{International Center for Quantum Materials, Peking University, Beijing 100871, China}
\affiliation{Collaborative Innovation Center of Quantum Matter, Beijing 100871, China}
\affiliation{Beijing Key Laboratory of Quantum Devices, Peking University, Beijing 100871, China}

\date{\today}

\begin{abstract}

  Topological orders in realistic systems are quite rare,
  raising one fundamental question of whether they might be harbored with
  hidden degrees of freedom rather than directly exposed in conventional spins or electrons.
  We here propose a set of hidden variables and their effective interactions
  in a bilayer kagom\'e antiferromagnet.
  By properly organizing four local states of the spin dimer into
  two pseudospin-$1/2$ variables ($\boldsymbol{\tau}$ and $\boldsymbol{\mu}$),
  we identify a regime where an effective Balents-Fisher-Girvin model emerges,
  realizing a $\mathbb{Z}_2$ topological order in the hidden $\boldsymbol{\tau}$ sector
  while the auxiliary $\boldsymbol{\mu}$ sector remains a trivial paramagnet.
  This hidden nature drastically alters the experimental manifestations.
  Physical spins map to composite objects, typically convolving fractionalized anyon
  continua with non-topological magnon excitations.
  Nevertheless, specific symmetry-selected channels, such as the longitudinal intralayer response,
  exclusively reveal the bare two-vison continuum without collective-mode dressing.
  Our findings demonstrate that the topological order could be buried in certain hidden and emergent
  degrees of freedom,
  pointing to a promising pathway for seeking elusive topological states
  disguised within hidden internal degrees of freedom.

\end{abstract}

\maketitle

\section{Introduction}

Topological order~\cite{wen_TopologicalOrdersRigid_1990} remains exceedingly scarce in nature.
Beyond the celebrated fractional quantum Hall effect~\cite{tsui_TwoDimensionalMagnetotransportExtreme_1982,laughlin_AnomalousQuantumHall_1983} with the recently observed fractional Chern insulators~\cite{cai_SignaturesFractionalQuantum_2023,zeng_ThermodynamicEvidenceFractional_2023,park_ObservationFractionallyQuantized_2023} largely representing lattice-based variants thereof,
the unambiguous realizations of topological order in physical systems are virtually non-existent.
From a mathematical and field-theoretic perspective,
one can engineer an abundance of exact lattice models~\cite{kitaev_FaulttolerantQuantumComputation_2003,levin_StringnetCondensationPhysical_2005};
yet these solvable constructions often inhabit idealized or physically unreachable Hilbert spaces,
rendering their relevance to realistic materials rather tenuous.
An exceptionally rare class of models that successfully bridges theoretical tractability
with real-world materials is exemplified by the Kitaev honeycomb model~\cite{kitaev_AnyonsExactlySolved_2006,jackeli_MottInsulatorsStrong_2009},
accompanied by a few less widely heralded yet profound paradigms
such as the Balents-Fisher-Girvin (BFG) model~\cite{balents_fractionalization_2002,zhao_z_2_2025} and the pyrochlore spin-$1/2$ XXZ model~\cite{hermele_PyrochlorePhotonsU1_2004}.
Setting these rare benchmark toy models aside,
realistic quantum materials inevitably confront us with complex,
bewildering experimental phenomena.
In such systems, the microscopic parent Hamiltonians are rarely known with certainty,
let alone analytically or numerically solvable.
Consequently, the fundamental connections linking ambiguous experimental anomalies,
underlying microscopic models, and any potential topological order remain thoroughly obscured.

\begin{figure}[b]
  \centering
  \includegraphics[width=0.90\columnwidth, clip]{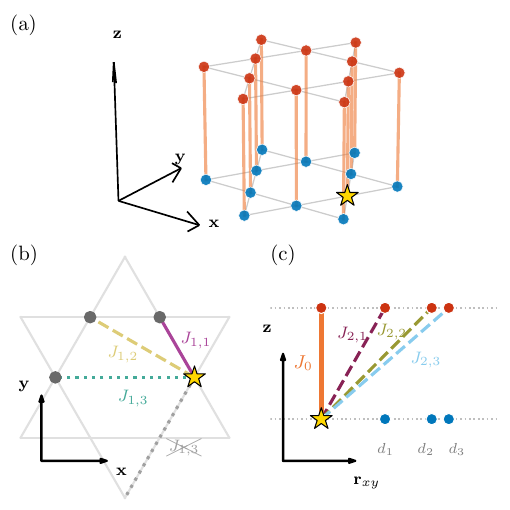}
  \caption{Bilayer kagom\'e lattice and couplings.
    (a) AA-stacked bilayer geometry.
    (b) Intralayer shells $J_{1,n}$; intrahexagon 3rd-neighbor couplings
    $J_{1,3}$ connect opposite vertices of the hexagon,
    while interhexagon 3rd-neighbor couplings (crossed out) are excluded.
    (c) The
    local exchange $J_0$ and nonlocal interlayer  $J_{2,n}$ at
  in-plane separations $d_n$. }
  \label{fig:bilayer_kagome_schematic}
\end{figure}

A major thrust in the quest for topological order has centered on quantum spin liquids (QSLs)~\cite{balents_SpinLiquidsFrustrated_2010}.
These strongly correlated magnetic phases evade conventional symmetry breaking
and instead exhibit long-range quantum entanglement, fractionalized excitations,
and emergent gauge fields~\cite{savary_QuantumSpinLiquids_2017}.
The mainstream efforts across frustrated magnetism, however, have almost exclusively
focused on conventional, bare spin degrees of freedom and their exchange interactions.
Under this standard paradigm, a central experimental diagnostic is the presence
of a broad spectral continuum~\cite{wen_ExperimentalIdentificationQuantum_2019}, such as spinon continua in frustrated magnets~\cite{han_FractionalizedExcitationsSpinliquid_2012,becker_DiagnosingFractionalizationSpin_2018}
or more specifically the Majorana continua in Kitaev materials~\cite{knolle_DynamicsTwoDimensionalQuantum_2014,banerjee_NeutronScatteringProximate_2017},
arising when a single microscopic spin flip fractionalizes
into multiple emergent quasiparticles.
Despite decades of intense theoretical and experimental pursuit,
indisputable confirmation of topological order in magnetic candidates
remains notoriously difficult~\cite{broholm_QuantumSpinLiquids_2020}. This long-standing impasse prompts a fundamental question.
Is the apparent scarcity of topological order an intrinsic material limitation?
Given the existence of fractional quantum Hall effects, we are more positive. Instead,
we raise a
different question.
Could genuine topological order secretly reside in hidden internal
degrees of freedom that evade direct exposure in conventional spin channels?

In this work, we provide a constructive answer to the above question
by proposing a concrete paradigm in an AA-stacked bilayer kagom\'{e} magnet.
In this architecture, each physical site hosts two vertically aligned spins,
one in each layer,
supplying a natural internal degree of freedom.
Utilizing an exact pseudospin representation~\cite{chen_emergent_2024},
we encode the four local spin configurations per site into two distinct pseudospin-1/2 variables,
\(\tau \) and \(\mu \).
In the regime of a dominant antiferromagnetic local interlayer Ising coupling,
the parallel alignment sector occurs at a very high energy,
leaving the hidden variable \(\tau \) as the active low-energy degree of freedom.
Remarkably, by properly tailoring the inter-site exchange profile,
the low-energy sector reduces exactly to the Balents-Fisher-Girvin (BFG) model~\cite{balents_fractionalization_2002,isakov_SpinLiquidPhaseSpin1_2006}.
Consequently, a \(\mathbb{Z}_{2}\) topological order, with emergent gauge fields, spinons, and visons,
is successfully realized within the hidden \(\tau \) sector,
while the auxiliary \(\mu \) sector remains a topologically trivial paramagnet.

This explicit realization provides a powerful demonstration of how hidden topological order
fundamentally alters experimental manifestations and alters conventional diagnostic rules.
By deriving a comprehensive physical operator dictionary that maps layer-resolved spin operators
to the combined \(\tau \)-\(\mu \) Hilbert space,
we reveal that physical spin probes typically capture composite objects.
A standard spin flip convolving the fractionalized anyon continua with the non-topological
magnon excitations yields broad, obscured profiles in the dynamic spin structure factor.
Our example demonstrates that this hidden nature does not mean the state is undetectable.
We identify specific symmetry-selected channels,
most notably the longitudinal same-layer response \(\mathcal{S}^{zz}_{\mathrm{u}\mathrm{u}}(\mathbf{k}, \omega)\), which satisfies a strict selection rule that completely strips away the magnon dressing,
exclusively exposing the bare two-vison continuum.
Our illustrative example ultimately establishes a new diagnostic principle for frustrated systems.
The apparent absence of clean fractionalized continua in conventional spectra does not rule out topological order, pointing to a promising uncharted pathway for seeking elusive topological states
disguised within hidden internal degrees of freedom.


The remainder of this paper is organized as follows. In Sec.~\ref{sec2},
we introduce the bilayer kagom\'e spin model and develop the $\tau$-$\mu$
pseudospin representation for the four local spin states at each in-plane site.
In Sec.~\ref{sec3}, we show how the hidden $\tau$ sector realizes a $\mathbb{Z}_2$ spin liquid
in the Balents-Fisher-Girvin limit.
In Sec.~\ref{sec4}, we describe the fractionalized spinon and vison excitations
alongside the auxiliary \(\mu \)-sector magnons.
In Sec.~\ref{sec5}, we establish the physical operator dictionary and derive the corresponding spectroscopic selection rules.
In Sec.~\ref{sec6}, we compute the dynamic spin structure factors
across various layer-resolved channels. Finally, in Sec.~\ref{sec7},
we discuss experimental implications and summarize our main conclusions

\section{Bilayer Kagom\'e Model and Pseudospin Representation}
\label{sec2}

To demonstrate how topological order can be sequestered within internal degrees of freedom,
we consider an AA-stacked bilayer kagom\'{e} spin-1/2 XXZ model.
At each spatial position \(i = (\mathbf{R}, a)\),
where \(\mathbf{R}\) denotes the triangular Bravais lattice vector
and \(a = 1, 2, 3\) labels the three kagom\'{e} sublattices,
the local Hilbert space is four-dimensional,
spanned by the vertically aligned upper-layer and lower-layer spins,
\(\mathbf{S}_{ui}\) and \(\mathbf{S}_{di}\).
The total microscopic Hamiltonian is given by
\begin{equation}
  H=H_{\mathrm{u}}+H_{\mathrm{d}}+H_0+H_{\mathrm{ud}}+H_Z ,
  \label{eq:microscopic_hamiltonian}
\end{equation}
where the individual layer contributions \(H_u, H_d\), the local vertical interlayer coupling \(H_{0}\), the nonlocal interlayer coupling \(H_{ud}\), and the Zeeman term \(H_{Z}\) are explicitly parameterized as,
\begin{align}
  H_{\mathrm{u}}
  &=
  \frac12\sum_{i\neq j}
  \qty[
    J^z_{1,ij}S^z_{\mathrm{u}i}S^z_{\mathrm{u}j}
    +J^\perp_{1,ij}
    \qty(S^x_{\mathrm{u}i}S^x_{\mathrm{u}j}+S^y_{\mathrm{u}i}S^y_{\mathrm{u}j})
  ],
  \nonumber\\
  H_{\mathrm{d}}
  &=
  \frac12\sum_{i\neq j}
  \qty[
    J^z_{1,ij}S^z_{\mathrm{d}i}S^z_{\mathrm{d}j}
    +J^\perp_{1,ij}
    \qty(S^x_{\mathrm{d}i}S^x_{\mathrm{d}j}+S^y_{\mathrm{d}i}S^y_{\mathrm{d}j})
  ],
  \nonumber\\
  H_0
  &=
  \sum_i
  \qty[
    J^z_0S^z_{\mathrm{u}i}S^z_{\mathrm{d}i}
    +J^\perp_0
    \qty(S^x_{\mathrm{u}i}S^x_{\mathrm{d}i}
    +S^y_{\mathrm{u}i}S^y_{\mathrm{d}i})
  ],
  \nonumber\\
  H_{\mathrm{ud}}
  &=
  \frac12\sum_{i\neq j}
  \Big[
    J^z_{2,ij}
    \qty(S^z_{\mathrm{u}i}S^z_{\mathrm{d}j}
    +S^z_{\mathrm{d}i}S^z_{\mathrm{u}j})
    \nonumber\\
    &\qquad
    +J^\perp_{2,ij}
    \qty(S^x_{\mathrm{u}i}S^x_{\mathrm{d}j}
    +S^x_{\mathrm{d}i}S^x_{\mathrm{u}j})
    \nonumber\\
    &\qquad
    +J^\perp_{2,ij}
    \qty(S^y_{\mathrm{u}i}S^y_{\mathrm{d}j}
    +S^y_{\mathrm{d}i}S^y_{\mathrm{u}j})
  \Big],
  \nonumber\\
  H_Z
  &=
  -h\sum_i
  \qty(S^z_{\mathrm{u}i}+S^z_{\mathrm{d}i}).
  \label{eq:bilayer_xxz_model}
\end{align}
Here, \(J_{0}\) specifies the local vertical dimer exchange, while \(J_{1}\) and \(J_{2}\) characterize the intralayer and nonlocal interlayer interactions across successive coordination shells \(n\), respectively.

To explicitly separate the hidden topological degree of freedom from the conventional magnetic sector,
we introduce a local change of variables that exactly maps the four-dimensional state space at each site
\(i\) onto two sets of pseudospin-1/2 operators, \(\vb*{\tau }_{i}\) and \(\vb*{\mu }_{i}\).
We define the quantized eigenstates of \(\tau_{i}^{z}\) and \(\mu_{i}^{z}\) via the physical spin configurations,
\begin{align}
  \ket{\tau_i^z=+\tfrac12,\mu_i^z=-\tfrac12}
  &=
  \ket{S^z_{\mathrm{u}i}=+\tfrac12,S^z_{\mathrm{d}i}=-\tfrac12},
  \nonumber\\
  \ket{\tau_i^z=-\tfrac12,\mu_i^z=-\tfrac12}
  &=
  \ket{S^z_{\mathrm{u}i}=-\tfrac12,S^z_{\mathrm{d}i}=+\tfrac12},
  \nonumber\\
  \ket{\tau_i^z=+\tfrac12,\mu_i^z=+\tfrac12}
  &=
  \ket{S^z_{\mathrm{u}i}=+\tfrac12,S^z_{\mathrm{d}i}=+\tfrac12},
  \nonumber\\
  \ket{\tau_i^z=-\tfrac12,\mu_i^z=+\tfrac12}
  &=
  \ket{S^z_{\mathrm{u}i}=-\tfrac12,S^z_{\mathrm{d}i}=-\tfrac12}.
  \label{eq:tau_mu_basis}
\end{align}
In this exact representation, \(\mu^z_i = -1/2\) selectively identifies the local interlayer antiferromagnetic (antiparallel) subspace, whereas \(\mu^z_i = +1/2\) identifies the ferromagnetic (parallel) subspace.
Within either sector, the active variable \(\tau _{i}^{z}\) acts as a hidden pseudospin
that tracks the orientation of the upper-layer spin. This yields a mathematically
exact operator dictionary across the entire Hilbert space, as listed in
Table~\ref{tab:physical_spin_tau_mu_mapping}.

\begin{table}[b]
  \caption{Exact local mapping from physical spin components to the
  pseudospin variables $\tau_i$ and $\mu_i$.}
  \label{tab:physical_spin_tau_mu_mapping}
  \begin{ruledtabular}
    \begin{tabular}{ccc}
      Component $\alpha$ & Upper layer $S^\alpha_{\mathrm{u}i}$
      & Lower layer $S^\alpha_{\mathrm{d}i}$ \\
      \hline
      $z$ & $\tau_i^z$ & $2\tau_i^z\mu_i^z$ \\
      $x$ & $2\tau_i^x\mu_i^x$ & $\mu_i^x$ \\
      $y$ & $2\tau_i^y\mu_i^x$ & $2\tau_i^z\mu_i^y$ \\
    \end{tabular}
  \end{ruledtabular}
\end{table}

We emphasize that the mapping in Table~\ref{tab:physical_spin_tau_mu_mapping}
constitutes an exact algebraic change of variables without any low-energy truncation.
Substituting the relations in Table~\ref{tab:physical_spin_tau_mu_mapping} into Eq.~\eqref{eq:microscopic_hamiltonian}
reformulates the microscopic model into a coupled \(\tau \)-\(\mu \) quantum system with
\begin{equation}
  H=H_0+H_Z+H_{\mathrm{Ising}}+H_\perp,
  \qquad
  H_\perp=H_{1,\perp}+H_{2,\perp},
  \label{eq:tau_mu_hamiltonian_decomposition}
\end{equation}
where the longitudinal and local transverse terms read
\begin{align}
  H_0
  &=
  \sum_i
  \qty[
    \frac{J_0^z}{2}\mu_i^z
    +J_0^\perp\tau_i^x\qty(\frac12-\mu_i^z)
  ],
  \nonumber\\
  H_Z
  &=
  -h\sum_i \tau_i^z\qty(1+2\mu_i^z),
  \nonumber\\
  H_{\mathrm{Ising}}
  &=
  \frac12\sum_{i\neq j}
  \qty[
    J^z_{1,ij}\qty(1+4\mu_i^z\mu_j^z)
    +2J^z_{2,ij}\qty(\mu_i^z+\mu_j^z)
  ]\tau_i^z\tau_j^z ,
  \label{eq:tau_mu_local_and_ising_terms}
\end{align}
and the nonlocal transverse couplings, which induce transitions
between different interlayer alignment sectors, are given by
\begin{align}
  H_{1,\perp}
  &=
  \frac12\sum_{i\neq j}J^\perp_{1,ij}
  \Bigg\{
    \mu_i^x\mu_j^x
    \qty[1+4\tau_i^x\tau_j^x+4\tau_i^y\tau_j^y]
    \nonumber\\
    &\qquad\qquad
    +4\tau_i^z\tau_j^z\mu_i^y\mu_j^y
  \Bigg\},
  \nonumber\\
  H_{2,\perp}
  &=
  \frac12\sum_{i\neq j}J^\perp_{2,ij}
  \Bigg[
    2\qty(\tau_i^x+\tau_j^x)\mu_i^x\mu_j^x
    \nonumber\\
    &\qquad
    +4\tau_i^y\tau_j^z\mu_i^x\mu_j^y
    +4\tau_i^z\tau_j^y\mu_i^y\mu_j^x
  \Bigg].
  \label{eq:tau_mu_transverse_terms}
\end{align}

\section{Strong-Interlayer Projection and the Construction of the BFG Model}
\label{sec3}

The mechanism for cloaking the topological order becomes operational in the strong-coupling limit,
where the local interlayer antiferromagnetic Ising exchange \(J^z_0 > 0\) dominates over all other energy scales. In this regime, the first term of \(H_{0}\) acts as a massive longitudinal field
that polarizes the auxiliary sector into the local antiparallel configuration
\(\mu^z_i = -1/2\) at every site.
This polarizes the \(\mu \) sector into a topologically trivial product state,
leaving a low-energy manifold separated from the parallel sector by a prominent bare gap of \(J^z_0/2\).
The Zeeman term simply vanishes in this regime.

Projecting the full Hamiltonian \(H\) onto this low-energy manifold at leading order amounts to freezing \(\mu^z_i \rightarrow -1/2\) and evaluating matrix elements within this gapped subspace. Under this projection, the Zeeman term and the nonlocal transverse terms \(H_{\perp }\) vanish entirely, contributing only via high-order virtual processes. The system collapses identically into an effective transverse-field Ising model (TFIM) governing only the hidden variables \(\tau \),
\begin{equation}
  \begin{aligned}
    H_{\mathrm{eff}}
    &=
    \frac12\sum_{i\neq j}
    V_{ij}\tau_i^z\tau_j^z
    +J_0^\perp\sum_i\tau_i^x
    -\frac{N J^z_0}{4},
  \end{aligned}
  \label{eq:effective_tau_tfim}
\end{equation}
where the effective inter-site exchange \(V_{ij}\) is uniquely dictated by the competition between intralayer and nonlocal interlayer interactions,
\begin{equation}
  V_{ij} = 2 \left( J^z_{1,ij} - J^z_{2,ij} \right).
\end{equation}
This TFIM arises from the intrinsic spin interactions rather than the external magnetic field.
Thus, it is an example of intrinsic TFIM~\cite{PhysRevResearch.1.033141,PhysRevResearch.2.043013}.

\begin{figure}[t]
  \centering
  \includegraphics[width=\columnwidth]{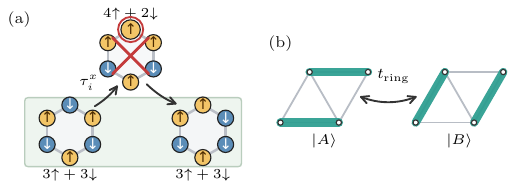}
  \caption{Construction of the BFG limit after the microscopic couplings of
    Fig.~\ref{fig:bilayer_kagome_schematic} have been projected to the
    effective $\tau$ model. (a) The equal-coupling condition
    $V_1=V_2=V_3$ penalizes nonzero $\tau^z$ charge on each kagom\'e hexagon. The
    shaded band denotes the low-energy three-up/three-down manifold. A single
    transverse-field flip creates constraint defects on two adjacent hexagons
    at an energy cost of order $V$. (b) Four virtual $\tau$ flips around a
    triangular-lattice rhombus return the system to the constrained manifold
    and resonate the two alternating link configurations within the
    three-dimer constrained Hilbert space, with matrix element
  $t_{\mathrm{ring}}=\mathcal{O}((J_0^\perp)^4/V^3)$.}
  \label{fig:bfg_construction}
\end{figure}

To drive this active \(\tau \) sector into a gapped \(\mathbb{Z}_{2}\) spin liquid,
we specialize to the coupling hierarchy introduced by Balents, Fisher, and Girvin (BFG).
We retain interactions up to the intrahexagon third-neighbor shell (\(n=1,2,3\))
while strictly excluding interhexagon third-neighbor paths.
By enforcing the BFG condition
\begin{equation}
  V_1=V_2=V_3\equiv V>0,
  \label{eq:bfg_condition}
\end{equation}
which maps onto a constraint on the difference profile \(2(J^z_{1,n} - J^z_{2,n}) = V\),
the effective Hamiltonian can be neatly recast into an unfrustrated sum of corner-sharing hexagonal clusters,
\begin{equation}
  H_{\mathrm{Ising}}^{\mathrm{BFG}}
  =
  \frac{V}{2}\sum_{\hexagon}
  \qty(\sum_{i\in \hexagon}\tau_i^z)^2
  -\frac{3V}{4}N_{\hexagon} ,
  \label{eq:bfg_hexagon_constraint}
\end{equation}
where $N_{\hexagon}$ is the number of kagom\'{e} hexagons. In the limit \(V \gg \vert{}J^\perp_0\vert{}\), the classical ground-state manifold is strictly defined by a local ice-like constraint on every single hexagon,
\begin{equation}
  \sum_{i\in\hexagon}\tau_i^z=0,
  \label{eq:bfg_three_up_three_down_constraint}
\end{equation}
requiring precisely three ``up'' and three ``down'' pseudospins per hexagon. Because a single transverse-field operator \(\tau _{i}^{x}\) acts as a defect creator that violates the ice rule on two adjacent hexagons at a severe energy penalty \(\mathcal{O}(V)\) [Fig.~\ref{fig:bfg_construction}(a)], the low-energy physics within the macroscopically degenerate BFG manifold is driven by a four-body ring-exchange resonance around the elementary rhombi of the lattice [Fig.~\ref{fig:bfg_construction}(b)]. This yields an effective dimer resonance amplitude,
\begin{equation}
  t_{\mathrm{ring}}=\mathcal{O}\qty(\frac{(J_0^\perp)^4}{V^3}).
  \label{eq:bfg_ring_exchange_scale}
\end{equation}
This maps exactly onto a gapped \(\mathbb{Z}_{2}\) quantum dimer model on the dual triangular lattice, ensuring that while the visible \(\mu \) sector is a completely trivial paramagnet, the hidden \(\tau \) sector hosts a fully functional, highly entangled \(\mathbb{Z}_{2}\) topological order.


\section{Fractionalized Excitations and Magnons}
\label{sec4}


The highly disparate roles of the two sectors manifest sharply in their elementary excitations. The hidden \(\tau \) sector hosts fractionalized, topological anyons, electric spinons and magnetic visons, whereas the auxiliary \(\mu \) sector produces conventional, non-topological collective modes (magnons).
We describe these
excitations using a gauge-fixed spinon mean field, a dual vison model, and a
spin-wave expansion for $\mu$, respectively. The operator mappings below
are exact within the stated gauge constraints, and the quadratic spectra are
approximate descriptions of the excitations.

\begin{figure}[t]
  \centering
  \includegraphics[width=0.8\columnwidth]{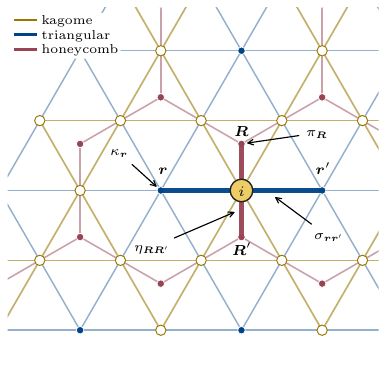}
  \caption{Lattice dictionary for the gauge and duality descriptions. Kagom\'e
    sites (yellow) are midpoints of triangular-lattice links (blue), which
    carry $\sigma$ gauge fields; triangular sites host spinons $\kappa$. The
    dual honeycomb sites and links (red) host the vison variables $\pi$ and
    $\eta$. The highlighted kagom\'e site $i$ lies at the midpoint of
  $\nnb{\rvec\rvec'}$ and is crossed by $\nnb{\vb*{R}\vb*{R}'}$.}
  \label{fig:lattice_duality_map}
\end{figure}

\begin{figure*}[t]
  \centering
  \includegraphics[width=0.96\textwidth]{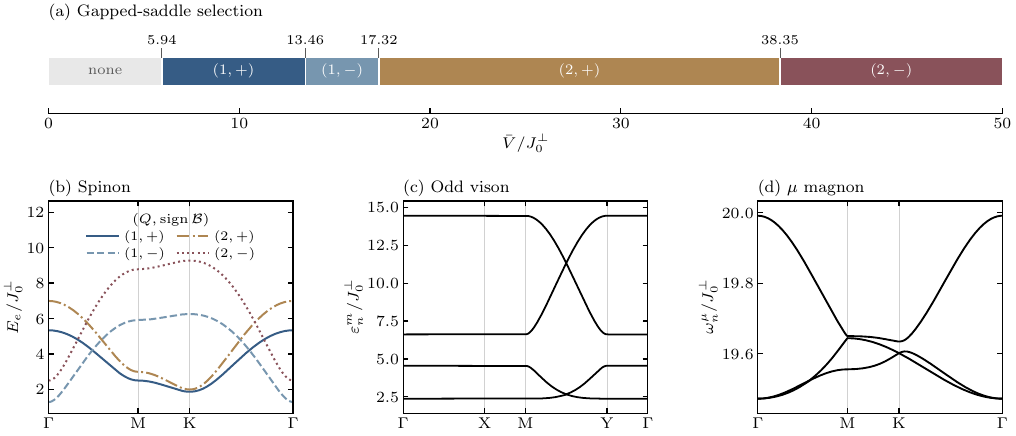}
  \caption{Gapped spinon saddle selection and representative excitation
    spectra. (a) Lowest constrained energy among uncondensed, gapped spinon
    saddles, labelled by $(Q,\operatorname{sign}\mathcal B)$.
    Gray denotes the absence of a gapped interior minimum in this ansatz.
    (b) Spinon bands at $\bar V/J_0^\perp=10,15,25,45$, in the sector order
    shown in (a), with solid, dashed, dash-dotted, and dotted lines,
    respectively. Each curve is the single spinon band of a different
    self-consistent solution, identified by the legend; the colors are
    shared with (a).
    We use $J_0^z/J_0^\perp=35$ and fix $\expval{\mu^z}=-1/2$.
    (c) Static vison-kernel eigenvalues for the odd background at
    $(\lambda_\pi,J_1^\pi,J_2^\pi)=(7,5,1)$.
    (d) Three auxiliary magnon bands for the effective coefficients
    $\epsilon_\mu=19.625$,
    $(t_1^\mu,t_2^\mu,t_3^\mu)=(0.055,0.033,0.0125)$, and
    $(d_1^\mu,d_2^\mu,d_3^\mu)=(0.095,0.057,0.0225)$ defined in
    Appendix~\ref{app:mu_spinwave}.
    All energies and effective coefficients are in units of $J_0^\perp$.
    The vison and magnon panels use independently specified effective
    parameters. Panel (a) compares gapped candidates only, without selecting
  against competing condensed states.}
  \label{fig:elementary_dispersions}
\end{figure*}

\subsection{Spinons}

We formalize the spinon matter fields by mapping the constrained \(\tau \) sector onto a \(\mathbb{Z}_{2}\) lattice gauge theory on the triangular lattice formed by the hexagon centers. Each kagomé site \(i\) is identified as a link \(\ell = \langle rr' \rangle\) connecting nearest-neighbor triangular sites. We identify the longitudinal variable as the link gauge field \(\sigma^x_{rr'} \equiv 2\tau^z_i\), and introduce a hardcore bosonic matter field \(b_{r}\) (with \(\kappa^x_r = -1 + 2b^\dagger_r b_r\)) at the vertices to represent the spinons. The physical states are explicitly governed by the Gauss-law constraint,
\begin{equation}
  \kappa^x_{\rvec}
  =
  \prod_{\rvec'\in \hexstar \rvec}
  \sigma^x_{\rvec\rvec'},
  \label{eq:gauss_law}
\end{equation}
where the product runs over the six links touching $\rvec$. With this
constraint, the transverse $\tau$ operator is represented as
$2\tau_i^x=\sigma^z_{\rvec\rvec'}\kappa^z_{\rvec}\kappa^z_{\rvec'}$.
In the three-up-three-down manifold,
\begin{equation}
  \prod_{\rvec'\in\hexstar\rvec}\sigma^x_{\rvec\rvec'}=-1,
  \label{eq:bfg_odd_background}
\end{equation}
so the BFG liquid has an odd background gauge charge. A local $\tau^x$
flip changes this parity at the two adjacent hexagon centers, creating a
spinon pair relative to that background. The parity condition retains less
information than the three-dimer constraint, which also fixes the
number of occupied links at each triangular site. The odd background fixes
the flux experienced by visons, as discussed in Sec.~\ref{sec:visons}.

To formulate the spinon mean-field theory, we represent the link and matter
variables by hardcore bosons,
\begin{equation}
  \begin{aligned}
    \sigma^x_\ell&=1-2B^\dagger_\ell B_\ell,\\
    \sigma^z_\ell&=B^\dagger_\ell+B_\ell,\\
    \kappa^z_{\rvec}&=b^\dagger_{\rvec}+b_{\rvec},\\
    \kappa^x_{\rvec}&=-1+2b^\dagger_{\rvec}b_{\rvec}.
  \end{aligned}
\end{equation}
Here the matter-boson vacuum is defined relative to the odd background.
For a uniform real mean-field ansatz in a fixed gauge, we replace the link operator by
$\mathcal{B}\equiv\expval*{B_{\rvec\rvec'}}=\expval*{B^\dagger_{\rvec\rvec'}}$
and fix the auxiliary background at $\expval{\mu^z}=-1/2$.
Relaxing the matter hardcore constraint to a
quadratic boson approximation gives
\begin{equation}
  \begin{aligned}
    H_e={}&J_e\sum_{\nnb{\rvec\rvec'}}
    \qty(b^\dagger_{\rvec}+b_{\rvec})
    \qty(b^\dagger_{\rvec'}+b_{\rvec'})\\
    &+\lambda\sum_{\rvec}b^\dagger_{\rvec}b_{\rvec}+E_c,\\
    J_e={}&J_0^\perp\mathcal B.
  \end{aligned}
  \label{eq:spinon_mf_hamiltonian}
\end{equation}
The hopping and pair-creation amplitudes have the common coefficient $J_e$.
The parameter $\lambda$ sets the matter-boson energy and is fixed by an
average number constraint in the saddle-point approximation described in
Appendix~\ref{app:spinon_mf}. We define
$  \gamma_{\kvec}
=
\sum_{\vb*{\delta}}
\ee^{-\ii\kvec\cdot\vb*{\delta}}
$,
where $\vb*{\delta}$ runs over the six nearest-neighbor vectors. The Bogoliubov
diagonalization gives a single spinon branch
\begin{equation}
  E_e(\kvec)
  =
  \sqrt{\lambda\qty(\lambda+2J_e\gamma_{\kvec})}.
  \label{eq:spinon_dispersion_main}
\end{equation}
Stability of this quadratic Hamiltonian requires $\lambda>0$ and
$\lambda+2J_e\gamma_{\kvec}>0$ throughout the Brillouin zone. For the uniform
ansatz, the band minimum is at $K$ for $J_e>0$ and at $\Gamma$ for $J_e<0$.
A gapped spinon saddle is a consistency condition for the assumed liquid;
deconfinement also requires the vison sector to remain gapped.

We compare number-constrained spinon saddles with $Q=1,2$ and
$\mathcal B\gtrless0$. Here $Q$ is the integer target in the softened
constraint $\expval{b^\dagger b}+6\mathcal B^2=Q$, and its parity labels
the chosen background. This approximation does not impose the three-dimer
constraint, even when $Q$ is odd. The even background, included here for
comparison, can be favored by a Zeeman term acting on $\tau^z$ that selects
an even number of dimers at each triangular-lattice site. The scan variable is
\begin{equation}
  \begin{aligned}
    \bar V&=2\qty(\tilde V_1^z-\tilde V_2^z),\\
    \tilde V_a^z&=2J^z_{a,1}+2J^z_{a,2}+J^z_{a,3},\qquad a=1,2.
  \end{aligned}
  \label{eq:spinon_vbar_definition}
\end{equation}
At $V_n=V$, this gives $\bar V=5V$. The energy functional
and stationarity equations used for this comparison are given in
Appendix~\ref{app:spinon_mf}.

The spinon gap is $\Delta_e=\min_{\kvec}E_e(\kvec)$. We restrict the
candidate set to uncondensed saddles with $\Delta_e>0$.
For each $Q$ and sign of $\mathcal B$,
we enforce $\partial f/\partial\lambda=0$ and minimize the energy along
the resulting constraint curve. Brillouin-zone averages are evaluated in
the thermodynamic limit. Among the admissible gapped minima, the selected
sectors are successively $(Q,\operatorname{sign}\mathcal B)=(1,+),(1,-),
(2,+),(2,-)$. Gapless and condensed solutions are
excluded from this comparison. The boundaries thus mark the appearance of
new gapped candidates, rather than energy crossings with ordered phases.
Fig.~\ref{fig:elementary_dispersions}(a,b) illustrates the selected sectors
and their representative spinon dispersions; the onset values are listed
in Table~\ref{tab:spinon_gapped_onsets}.

\subsection{Visons}
\label{sec:visons}

A vison changes the $\mathbb Z_2$ magnetic flux through a triangular
plaquette. The plaquette centers form the dual honeycomb lattice, whose
sites coincide with the kagom\'e triangle centers. We introduce Pauli
operators $\pi_{\vb*{R}}$ for the dual matter and
$\eta_{\vb*{R}\vb*{R}'}$ for the dual gauge field. Electric--magnetic
duality maps the original magnetic flux to $\pi^x$ and the original
electric charge to a dual hexagon flux~\cite{zhao_z_2_2025}.
The complete operator dictionary, both Gauss laws, and the transformed
Hamiltonian are given in Appendix~\ref{app:vison_duality}. The relation
relevant for the spectroscopy is
\begin{equation}
  \sigma^x_{\rvec\rvec'}
  =
  \eta^z_{\vb*{R}\vb*{R}'}
  \pi^z_{\vb*{R}}\pi^z_{\vb*{R}'},
  \label{eq:vison_duality}
\end{equation}
where the honeycomb link $\nnb{\vb*{R}\vb*{R}'}$ crosses the
triangular link $\nnb{\rvec\rvec'}$. The right-hand side is a
gauge-invariant vison-pair operator.
It changes $\pi^x$ at the two endpoints,
and the link factor $\eta^z$ supplies the required gauge dressing.
Thus the
longitudinal $\tau^z\sim \sigma^x/2$ operator creates two magnetic-flux excitations
relative
to the chosen background.

We describe the vison fluctuations by approximating the spinon charges as a
static background. In a fixed dual $\mathbb{Z}_2$ gauge, we replace the gauge field by a
prescribed mean-field pattern $\bar{\eta}^z_{\vb*{R}\vb*{R}'}$ consistent with the
Gauss law. We use a soft-spin
approximation, which replaces the two-level Pauli component $\pi^z_{\vb*{R}}$
by a continuous real field $\phi_{\vb*{R}}$~\cite{huh_VisonStatesConfinement_2011}.
This relaxes the exact identity $(\pi^z_{\vb*{R}})^2=1$ and permits
amplitude fluctuations of $\phi_{\vb*{R}}$. The field $\phi$ describes
vison excitations. Under these approximations, the vison Hamiltonian becomes quadratic in the soft field,
\begin{align}
  H_m
  &=
  J^\pi_1
  \sum_{\nnb{\vb*{R}\vb*{R}'}}
  \bar{\eta}^z_{\vb*{R}\vb*{R}'}
  \phi_{\vb*{R}}\phi_{\vb*{R}'}
  +
  J^\pi_2
  \sum_{\langle\!\langle \vb*{R}\vb*{R}'\rangle\!\rangle}
  \bar{\eta}^z_{\vb*{R}\vb*{R}''}
  \nonumber\\
  &\quad\times
  \bar{\eta}^z_{\vb*{R}''\vb*{R}'}
  \phi_{\vb*{R}}\phi_{\vb*{R}'}
  +\lambda_\pi\sum_{\vb*{R}}\qty(\phi_{\vb*{R}}^2-1)
  +E_m^{(0)} .
  \label{eq:vison_soft_hamiltonian_main}
\end{align}
Here $\vb*{R}''$ is the common neighbor of the next-nearest-neighbor pair.
The microscopic origin and limitations of this truncation are discussed in
Appendix~\ref{app:vison_duality}. In particular, dual links sharing a honeycomb
site yield a quadratic term because $(\pi^z_{\vb*{R}''})^2=1$, whereas
disjoint links retain four-vison interactions. The latter also occur at the
BFG point and cannot be discarded as weak further-neighbor corrections.
Accordingly, $J_1^\pi,J_2^\pi$, and $\lambda_\pi$ are treated as the
phenomenological parameters of the quadratic vison theory. The microscopic
symmetry restrictions on this parametrization are given in
Appendix~\ref{app:vison_duality}. The odd constraint of the present model
fixes the background flux to
\begin{equation}
  W=\prod_{\ell\in\hexagon}\bar{\eta}^z_\ell
  =-1.
  \label{eq:odd_vison_flux}
\end{equation}
Here $\hexagon$ labels a hexagonal plaquette of the dual honeycomb lattice around one
triangular site.

For the odd BFG background, $W=-1$ follows directly from
Eq.~\eqref{eq:bfg_odd_background}. A convenient gauge doubles the honeycomb
unit cell, and the vison kernel becomes a four-band matrix in the magnetic
Brillouin zone. The enlarged unit cell reflects fractionalized translation
symmetry of the vison, not symmetry breaking of the spin-liquid ground state.
The gauge convention and quadratic kernel are given in
Appendix~\ref{app:vison_duality}.

Diagonalizing the quadratic kernel gives the vison bands
$\varepsilon_n^m(\kvec)$.
The four odd-background bands are shown in
Fig.~\ref{fig:elementary_dispersions}(c).

\subsection{Auxiliary magnons}

The $\mu$ sector supplies ordinary non-topological collective modes. The
nonlocal transverse term $H_\perp$, absent from the leading projected
Hamiltonian but present in the exact model, gives these excitations a
dispersion. We therefore expand about the polarized state $\mu_i^z=-1/2$,
\begin{equation}
  \mu_i^z=a_i^\dagger a_i-\frac12,\qquad
  \mu_i^+\simeq a_i^\dagger,\qquad
  \mu_i^-\simeq a_i .
  \label{eq:mu_hp_main}
\end{equation}
At zero field, we replace the $\tau$ operators by their equal-time
expectation values in the symmetry-preserving spin-liquid ground state.
Thus $\expval{\tau^z}=0$, while the single-site transverse expectation
$\expval{\tau^x}$ and bond correlators are retained. In the strictly
constrained manifold, acting with $\tau_i^z$ on
$\sum_{j\in\hexagon}\tau_j^z=0$ gives
\begin{equation}
  \sum_{j\in\hexagon,\,j\neq i}\expval{\tau_i^z\tau_j^z}=-\frac14
  \label{eq:hexagon_sum_rule}
\end{equation}
for each of the two hexagons containing site $i$. This fixes the sum of
the five bond correlators, not each correlator separately. A uniform value
$-1/20$ per bond is one possible approximation to the sum rule. Virtual
constraint violations give corrections of order $(J_0^\perp/V)^2$.

Changing $\mu_i^z$ from $-1/2$ to $+1/2$ removes the Ising interactions
between site $i$ and its ten intrahexagon neighbors. In a frozen $\tau$
background, the on-site coefficient is therefore
\begin{equation}
  \begin{aligned}
    \epsilon_\mu
    &=\frac{J_0^z}{2}-\sum_{j\neq i}V_{ij}
    \expval{\tau_i^z\tau_j^z}-J_0^\perp\expval{\tau^x},\\
    &=\frac{J_0^z}{2}+\frac{V}{2}
    +\mathcal O\qty(\frac{(J_0^\perp)^2}{V})\quad\text{at the BFG point}.
  \end{aligned}
  \label{eq:mu_gap_formula}
\end{equation}
The $V/2$ term follows from summing Eq.~\eqref{eq:hexagon_sum_rule} over
both hexagons. A single virtual $\tau$ flip has energy cost $V$ and matrix
element $J_0^\perp/2$, giving
\begin{equation}
  \begin{aligned}
    \frac{E^{(2)}}{N}&=-\frac{(J_0^\perp)^2}{4V},\\
    \expval{\tau^x}&=-\frac{J_0^\perp}{2V}
    +\mathcal O\qty(\frac{(J_0^\perp)^3}{V^3}).
  \end{aligned}
  \label{eq:tau_transverse_expectation}
\end{equation}
The transverse contribution to $\epsilon_\mu$ is positive. Corrections to
the bond correlators enter at the same order and must also be retained for
a quantitative expansion through $(J_0^\perp)^2/V$. Eq.~\eqref{eq:mu_gap_formula}
is an on-site estimate within the factorized background; the actual band gap
includes dispersion and relaxation of the surrounding $\tau$ state.

The transverse intralayer and nonlocal
interlayer couplings generate both normal and anomalous bilinears on the
three-site kagom\'e unit cell. To formulate the quadratic Hamiltonian,
we introduce the Nambu spinor
\begin{equation}
  \Phi_{\kvec}=
  \qty(a_{\kvec,1},a_{\kvec,2},a_{\kvec,3},
  a^\dagger_{-\kvec,1},a^\dagger_{-\kvec,2},a^\dagger_{-\kvec,3})^T,
  \label{eq:mu_nambu_spinor}
\end{equation}
where $a_{\kvec,s}$ annihilates an auxiliary magnon on sublattice
$s=1,2,3$. The quadratic Hamiltonian is a bosonic BdG problem,
\begin{equation}
  H_\mu^{(2)}
  =
  \frac12\sum_{\kvec}
  \Phi^\dagger_{\kvec}
  \mathcal{H}_\mu(\kvec)
  \Phi_{\kvec}
  +E_\mu^{(0)},
  \label{eq:mu_bdg_main}
\end{equation}
with the matrix $\mathcal H_\mu(\kvec)$ specified in
Appendix~\ref{app:mu_spinwave}. The positive bosonic BdG energies define
three branches $\omega_n^\mu(\kvec)$, with the gap
$\Delta_\mu=\min_{\kvec,n}\omega_n^\mu(\kvec)$. In the strong-interlayer
regime their bandwidth is small compared with the local scale $J_0^z/2$.
These modes carry no topological charge. Their role in dressing the physical
spin response follows from the operator dictionary developed in Sec.~\ref{sec5}.

\section{Physical Operator Dictionary and Spectroscopic Selection Rules}
\label{sec5}

The response measured by a microscopic spin operator is determined by how that
operator resolves into the hidden topological sector and the auxiliary $\mu$
sector. We use the Pauli-normalized notation
$\nu_i^\alpha=2\mu_i^\alpha$ for the auxiliary variable, and similarly write
$2\tau_i^\alpha$ for local operators in the BFG sector. For a kagom\'e site $i$
lying at the midpoint of a triangular link $\nnb{\rvec_1\rvec_2}$ and a dual
honeycomb link
$\nnb{\vb*{R}_1\vb*{R}_2}$, define
\begin{align}
  \mathcal{O}^e_i
  &\equiv
  2\tau_i^x
  =
  \sigma^z_{\rvec_1\rvec_2}
  \kappa^z_{\rvec_1}\kappa^z_{\rvec_2}
  =
  \eta^x_{\vb*{R}_1\vb*{R}_2},
  \nonumber\\
  \mathcal{O}^m_i
  &\equiv
  2\tau_i^z
  =
  \sigma^x_{\rvec_1\rvec_2}
  =
  \eta^z_{\vb*{R}_1\vb*{R}_2}
  \pi^z_{\vb*{R}_1}\pi^z_{\vb*{R}_2},
  \nonumber\\
  \mathcal{O}^\epsilon_i
  &\equiv
  \ii\mathcal{O}^e_i\mathcal{O}^m_i
  =
  2\tau_i^y .
  \label{eq:topological_local_operators}
\end{align}
Here the superscripts label the topological sectors. $\mathcal{O}^e$ and
$\mathcal{O}^m$ create spinon and vison pairs, respectively, while
$\mathcal{O}^\epsilon$ probes the composite $\epsilon=e\times m$ channel.
Because these local operators are strictly gauge invariant, they always
excite anyon pairs or multi-anyon continua rather than isolated anyons.

Table~\ref{tab:operator_dictionary} summarizes the notation used to connect
the microscopic spin variables, the auxiliary $\mu$ variables, and the
emergent gauge-theory variables.

\begin{table}[t]
  \caption{Operator dictionary. The kagom\'e site $i$ is the midpoint of
    the triangular link $\nnb{\rvec_1\rvec_2}$ and is crossed by the honeycomb
  link $\nnb{\vb*{R}_1\vb*{R}_2}$.}
  \label{tab:operator_dictionary}
  \begin{ruledtabular}
    \begin{tabular}{p{0.18\columnwidth}
        p{0.36\columnwidth}
      p{0.40\columnwidth}}
      Symbol & Definition & Meaning \\
      \hline
      $S^\alpha_{\mathrm{u}i},S^\alpha_{\mathrm{d}i}$ &
      Table~\ref{tab:physical_spin_tau_mu_mapping} &
      microscopic spin \\
      $\tau_i$ &
      Eq.~\eqref{eq:tau_mu_basis} &
      hidden BFG variable \\
      $\mu_i,\nu_i$ &
      $\nu_i^\alpha=2\mu_i^\alpha$, Eq.~\eqref{eq:tau_mu_basis} &
      auxiliary alignment variable \\
      $\sigma_{\rvec_1\rvec_2}$ &
      $2\tau_i^z=\sigma^x_{\rvec_1\rvec_2}$ &
      gauge field on a triangular link \\
      $\kappa_{\rvec}$ &
      $2\tau_i^x=\sigma^z_{\rvec_1\rvec_2}
      \kappa^z_{\rvec_1}\kappa^z_{\rvec_2}$ &
      spinon matter field \\
      $\eta_{\vb*{R}_1\vb*{R}_2},\pi_{\vb*{R}}$ &
      $2\tau_i^z=
      \eta^z_{\vb*{R}_1\vb*{R}_2}
      \pi^z_{\vb*{R}_1}\pi^z_{\vb*{R}_2}$ &
      dual vison variables \\
      $\mathcal{O}^e_i$ &
      $2\tau_i^x$ &
      two-spinon probe \\
      $\mathcal{O}^m_i$ &
      $2\tau_i^z$ &
      two-vison probe \\
      $\mathcal{O}^\epsilon_i$ &
      $\ii\mathcal{O}^e_i\mathcal{O}^m_i=2\tau_i^y$ &
      $\epsilon=e\times m$ probe \\
    \end{tabular}
  \end{ruledtabular}
\end{table}

The physical spin operators follow by substituting these gauge-invariant
objects into the exact local $\tau$-$\mu$ dictionary of
Table~\ref{tab:physical_spin_tau_mu_mapping},
\begin{equation}
  \begin{aligned}
    S^z_{\mathrm{u}i}
    &=
    \frac12\mathcal{O}^m_i,
    &\qquad
    S^z_{\mathrm{d}i}
    &=
    \frac12\mathcal{O}^m_i\nu_i^z,
    \\
    S^x_{\mathrm{u}i}
    &=
    \frac12\mathcal{O}^e_i\nu_i^x,
    &\qquad
    S^x_{\mathrm{d}i}
    &=
    \frac12\nu_i^x,
    \\
    S^y_{\mathrm{u}i}
    &=
    \frac12\mathcal{O}^\epsilon_i\nu_i^x,
    &\qquad
    S^y_{\mathrm{d}i}
    &=
    \frac12\mathcal{O}^m_i\nu_i^y.
  \end{aligned}
  \label{eq:physical_operator_dictionary_main}
\end{equation}
Remarkably, the upper-layer longitudinal spin $S^z_{\mathrm{u}i}$ is
proportional to $\tau_i^z$ alone, providing a direct, undressed probe of the
vison-pair operator $\mathcal{O}^m_i$. By contrast, the lower-layer
component $S^z_{\mathrm{d}i}$ is coupled to $\nu_i^z$. Expanding around the
polarized background as $\nu_i^z=\nu_0+\delta\nu_i^z$ (with
$\nu_0=2\expval{\mu_i^z}\simeq -1$ at large $J_0^z$), one sees that its
static part still accesses the bare two-vison continuum, whereas the
fluctuation $\delta\nu_i^z$ introduces inelastic dressing from the $\mu$
sector. Meanwhile, the transverse spin components involve $\nu_i^{x,y}$ and
thus act as purely composite operators at leading order, except for
$S^x_{\mathrm{d}i}$ which directly measures the auxiliary $\mu$ magnons.

We denote the layer-resolved dynamic spin structure factor by
$\mathcal S^{\alpha\beta}_{ab}(\kvec,\omega)$, with
$a,b=\mathrm{u},\mathrm{d}$ specifying the layers and $\alpha,\beta$ the spin
components. The longitudinal same-layer response of the upper layer is thus
$\mathcal S^{zz}_{\mathrm{u}\mathrm{u}}$, whose spectral representation is
given in Sec.~\ref{sec6}.

Thus the special role of $\mathcal S^{zz}_{\mathrm{u}\mathrm{u}}$ is not that it is the
only longitudinal channel with a bare vison component. Rather, it is the
simplest layer-resolved channel because its local operator is exactly
proportional to $\mathcal{O}^m$. In a physical bilayer probe the natural
diagnostics are the layer-even and layer-odd longitudinal combinations, which
are fixed within the antiparallel manifold by the exact identities
$P_\mu(S^z_{\mathrm{u}}+S^z_{\mathrm{d}})P_\mu
=\frac12 P_\mu\,\mathcal{O}^m(1+\nu^z)\,P_\mu=0$ and
$P_\mu(S^z_{\mathrm{u}}-S^z_{\mathrm{d}})P_\mu=\mathcal{O}^m$. The low-energy
longitudinal response resides entirely in the layer-odd combination, with
four times the intensity of the single-layer channel, while the layer-even
combination carries only $\mu$-dressed weight above $2\Delta_\mu$. This does
not mean that the bare channel must carry
the largest spectral weight in a real material. It means that its operator
content has the fewest auxiliary-sector ambiguities.

\begin{figure*}[t]
  \centering
  \includegraphics[width=0.95\textwidth]{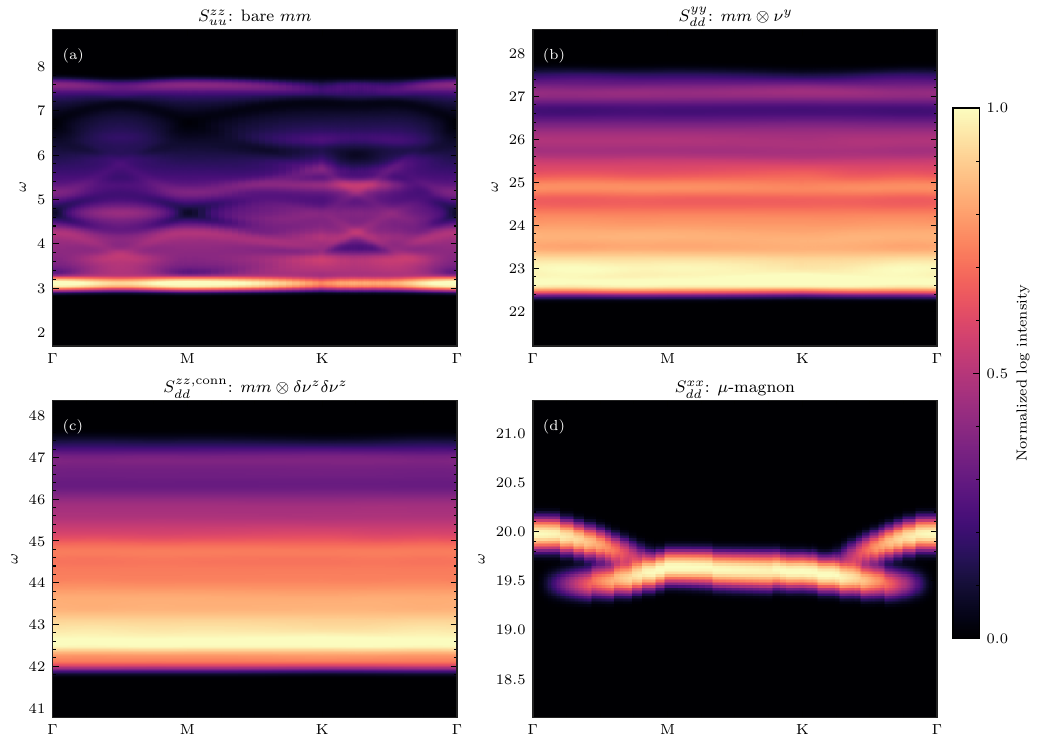}
  \caption{Representative layer-resolved diagonal dynamical structure factors
    along $\Gamma$-M-K-$\Gamma$. The panels contrast the clean diagnostic
    $\mathcal{S}^{zz}_{\mathrm{u}\mathrm{u}}=\mathcal{S}_{mm}/4$ with the
    connected fluctuation-dressed lower-layer longitudinal response and
    channels in which the same vison continuum is dressed by auxiliary
    $\mu$-sector excitations, including
    $\mathcal{S}^{yy}_{\mathrm{d}\mathrm{d}}
    \simeq\mathcal{S}_{mm}*\mathcal{S}_{\nu^y\nu^y}/4$ and
    the connected part of
    $\mathcal{S}^{zz}_{\mathrm{d}\mathrm{d}}$,
    $\mathcal{S}_{mm}*
    \mathcal{S}^{\mathrm{conn}}_{\delta\nu^z\delta\nu^z}/4$, together with the
    topologically trivial control channel
    $\mathcal{S}^{xx}_{\mathrm{d}\mathrm{d}}=\mathcal{S}_{\nu^x\nu^x}/4$.
    The full $\mathcal{S}^{zz}_{\mathrm{d}\mathrm{d}}$ contains the static
    bare-vison term $\nu_0^2\mathcal{S}_{mm}/4$ in addition to the connected
    contribution.
  Intensities are normalized within each panel on a logarithmic scale.}
  \label{fig:dsf_channels}
\end{figure*}

\section{Dynamic Spin Structure Factors}
\label{sec6}

At zero temperature, the dynamic spin structure factor is given by the
Lehmann representation
\begin{equation}
  \begin{aligned}
    \mathcal S^{\alpha\beta}_{ab}(\kvec,\omega)
    ={}&\sum_n
    \bra{0}S^\alpha_{a,\kvec}\ket{n}
    \bra{n}S^\beta_{b,-\kvec}\ket{0}\\
    &\times\delta\qty[\omega-(E_n-E_0)],
  \end{aligned}
  \label{eq:structure_factor_definition}
\end{equation}
where $S^\alpha_{a,\kvec}=N^{-1/2}\sum_i\ee^{-\ii\kvec\cdot\vb*{x}_i}S^\alpha_{a i}$,
and $\ket{n}$ denotes an eigenstate with energy $E_n$ ($\ket{0}$ being the
ground state). Throughout this section, momenta are defined in the
kagom\'e Brillouin zone unless specified otherwise.

Using the operator dictionary in Eq.~\eqref{eq:physical_operator_dictionary_main},
we evaluate the structure factors by decoupling the hidden topological
sector from the auxiliary $\nu$ sector at leading order. A composite local
operator $A_i=B_iC_i$ then yields a momentum-energy convolution of the two
sectors,
\begin{equation}
  \mathcal{S}_{A}(\kvec,\omega)
  \simeq
  \frac{1}{N}
  \sum_{\vb*{q}}
  \int\dd\Omega\,
  \mathcal{S}_{B}(\vb*{q},\Omega)
  \mathcal{S}_{C}(\kvec-\vb*{q},\omega-\Omega).
  \label{eq:main_convolution}
\end{equation}
While residual inter-sector interactions beyond this leading factorization
may redistribute spectral weight near the thresholds, they do not alter the
kinematic onset or the fundamental distinction between bare and composite
channels.

The elementary topological building blocks are the two-spinon, two-vison, and
$\epsilon$-sector spectral functions. In the spinon channel,
\begin{align}
  \mathcal{S}_{ee}(\kvec,\omega)
  &=
  \frac{\abs{A_e}^2}{N}
  \sum_{\vb*{q}}
  \abs{F_e(\kvec,\vb*{q})}^2
  \ee^{-2\theta_{\vb*{q}}}
  \ee^{-2\theta_{\kvec-\vb*{q}}}
  \nonumber\\
  &\quad\times
  \delta\qty(
    \omega-E_e(\vb*{q})-E_e(\kvec-\vb*{q})
  ),
  \label{eq:spinon_continuum_main}
\end{align}
where $F_e$ is a kagom\'e bond form factor and $\theta_{\kvec}$ is the
Bogoliubov angle of Eq.~\eqref{eq:spinon_mf_hamiltonian}. The lower edge is
$\min_{\vb*{q}}\qty[E_e(\vb*{q})+E_e(\kvec-\vb*{q})]$.

For visons, the generic two-particle response is
\begin{align}
  \mathcal{S}_{mm}(\kvec,\omega)
  &=
  \frac{1}{N}
  \sum_{\vb*{q}}\sum_{n,n'}
  \abs{\mathcal{M}^m_{nn'}(\kvec,\vb*{q})}^2
  \nonumber\\
  &\quad\times
  \delta\qty(
    \omega-\omega^m_n(\vb*{q})
    -\omega^m_{n'}(\kvec-\vb*{q})
  ),
  \label{eq:vison_continuum_main}
\end{align}
where the matrix element contains the dual gauge background and honeycomb
sublattice form factors. The band indices run over the four odd-sector
vison bands. The kinematic two-vison threshold is
\begin{equation}
  \omega^m_{\mathrm{low}}(\kvec)
  =\min_{\vb*{q},n,n'}\qty[
  \omega^m_n(\vb*{q})+\omega^m_{n'}(\kvec-\vb*{q})],
  \label{eq:two_vison_threshold}
\end{equation}
where $\omega_n^m$ denotes the vison dispersion (see Appendix~\ref{app:vison_duality}),
though form-factor zeros may suppress the spectral weight right at the
kinematic edge. Neglecting potential spinon-vison bound states, we
similarly approximate the $\epsilon$-sector response as a four-anyon
continuum formed by convolving the two-spinon and two-vison spectra,

\begin{equation}
  \mathcal{S}_{\varepsilon\varepsilon}(\kvec,\omega)
  \simeq
  \frac{C_\varepsilon}{N}
  \sum_{\vb*{q}}
  \int\dd\Omega\,
  \mathcal{S}_{ee}(\vb*{q},\Omega)
  \mathcal{S}_{mm}(\kvec-\vb*{q},\omega-\Omega).
  \label{eq:epsilon_continuum_main}
\end{equation}

The auxiliary sector contributes one- and two-magnon responses. In the
Holstein-Primakoff expansion about $\mu^z=-1/2$, the transverse operators
$\nu^x$ and $\nu^y$ are linear in $a^\dagger$ and $a$, whereas the connected
part of $\nu^z$ is bilinear. Thus $\nu^x$ and $\nu^y$ couple to a single
$\mu$ magnon,
\begin{equation}
  \mathcal{S}_{\nu^\alpha\nu^\alpha}(\kvec,\omega)
  =
  \sum_{n=1}^{3}
  \abs{\mathcal{M}^{\alpha}_{n}(\kvec)}^2
  \delta\qty(\omega-\omega^\mu_n(\kvec)),
  \qquad
  \alpha=x,y ,
  \label{eq:nu_transverse_response_main}
\end{equation}
while the connected part of $\nu^z$ starts as a two-magnon continuum,
\begin{align}
  \mathcal{S}^{\mathrm{conn}}_{\nu^z\nu^z}(\kvec,\omega)
  &=
  \frac{1}{N}
  \sum_{\vb*{q}}\sum_{n,n'=1}^{3}
  \abs{\mathcal{M}^{z}_{nn'}(\kvec,\vb*{q})}^2
  \nonumber\\
  &\quad\times
  \delta\qty(
    \omega-\omega^\mu_n(\vb*{q})
    -\omega^\mu_{n'}(\kvec-\vb*{q})
  ) .
  \label{eq:nu_z_response_main}
\end{align}
The full $\nu^z$ response also contains a static contribution from
$\nu_0=2\expval{\mu^z}$. In the longitudinal lower-layer channel this static
factor multiplies the vison operator and therefore contributes another copy of
the bare two-vison continuum; only the connected fluctuation
$\delta\nu^z=\nu^z-\nu_0$ produces the dressed inelastic continuum.

Using Eq.~\eqref{eq:physical_operator_dictionary_main}, the layer-resolved
channels become
\begin{align}
  \mathcal{S}^{zz}_{\mathrm{u}\mathrm{u}}
  &=
  \frac14\mathcal{S}_{mm},
  \nonumber\\
  \mathcal{S}^{zz}_{\mathrm{d}\mathrm{d}}
  &\simeq
  \frac14\qty[
    \nu_0^2\mathcal{S}_{mm}
    +
    \mathcal{S}_{mm}*
    \mathcal{S}^{\mathrm{conn}}_{\delta\nu^z\delta\nu^z}
  ],
  \nonumber\\
  \mathcal{S}^{zz}_{\mathrm{u}\mathrm{d}}
  &\simeq
  \frac{\nu_0}{4}\,\mathcal{S}_{mm},
  \nonumber\\
  \mathcal{S}^{xx}_{\mathrm{u}\mathrm{u}}
  &\simeq
  \frac14\mathcal{S}_{ee}*\mathcal{S}_{\nu^x\nu^x},
  \nonumber\\
  \mathcal{S}^{xx}_{\mathrm{d}\mathrm{d}}
  &=
  \frac14\mathcal{S}_{\nu^x\nu^x},
  \nonumber\\
  \mathcal{S}^{yy}_{\mathrm{u}\mathrm{u}}
  &\simeq
  \frac14\mathcal{S}_{\varepsilon\varepsilon}
  *\mathcal{S}_{\nu^x\nu^x},
  \nonumber\\
  \mathcal{S}^{yy}_{\mathrm{d}\mathrm{d}}
  &\simeq
  \frac14\mathcal{S}_{mm}*\mathcal{S}_{\nu^y\nu^y}.
  \label{eq:channel_formulas_main}
\end{align}
Here $*$ denotes the momentum-energy convolution in
Eq.~\eqref{eq:main_convolution}. Notably, $\mathcal{S}^{zz}_{\mathrm{u}\mathrm{u}}$
is completely free of auxiliary-sector convolutions. The interlayer
longitudinal correlator $\mathcal{S}^{zz}_{\mathrm{u}\mathrm{d}}$ also
remains undressed at this order because the linear fluctuation
$\expval{\delta\nu^z}$ vanishes in the factorized ground state, saturating
the on-site sum rule $\expval{S^z_{\mathrm{u}i}S^z_{\mathrm{d}i}}=\nu_0/4$
dictated by $(\mathcal{O}^m_i)^2=1$.

The clean bare-vison channel is therefore
$\mathcal{S}^{zz}_{\mathrm{u}\mathrm{u}}(\kvec,\omega)$. Its continuum
onset follows Eq.~\eqref{eq:two_vison_threshold}, and its intensity is governed
by the dual honeycomb form factor in Eq.~\eqref{eq:vison_continuum_main}. The
same topological continuum also appears in
$\mathcal{S}^{zz}_{\mathrm{d}\mathrm{d}}$ and
$\mathcal{S}^{yy}_{\mathrm{d}\mathrm{d}}$, but there it is folded with a
two-magnon $\nu^z$ response or a one-magnon $\nu^y$ response. These convolutions
raise and broaden spectral thresholds and can obscure the momentum structure
of the underlying visons.

The three longitudinal channels combine into the response measured at
interlayer momentum transfer $k_z$ for two layers separated by $d$, with
$\mathcal{S}^{zz}_{\mathrm{d}\mathrm{u}}=\mathcal{S}^{zz}_{\mathrm{u}\mathrm{d}}$
by layer symmetry and hermiticity,
\begin{align}
  \mathcal{S}^{zz}_{\mathrm{tot}}(\kvec,k_z,\omega)
  &=
  \mathcal{S}^{zz}_{\mathrm{u}\mathrm{u}}
  +\mathcal{S}^{zz}_{\mathrm{d}\mathrm{d}}
  +2\cos(k_z d)\,\mathcal{S}^{zz}_{\mathrm{u}\mathrm{d}}
  \nonumber\\
  &\simeq
  \frac{1+\nu_0^2+2\nu_0\cos(k_z d)}{4}\,
  \mathcal{S}_{mm}
  \nonumber\\
  &\quad
  +\frac14\,
  \mathcal{S}_{mm}*\mathcal{S}^{\mathrm{conn}}_{\delta\nu^z\delta\nu^z}.
  \label{eq:bilayer_kz_response}
\end{align}
The dressed term begins only $2\Delta_\mu$ above the bare two-vison
threshold, so in the entire low-energy window
Eq.~\eqref{eq:bilayer_kz_response} reduces, for $\nu_0\to-1$, to
\begin{equation*}
  \mathcal{S}^{zz}_{\mathrm{tot}}
  \simeq
  \sin^2\!\qty(k_z d/2)\,\mathcal{S}_{mm}(\kvec,\omega).
\end{equation*}
Thus the bare two-vison continuum has this interlayer modulation and vanishes
at $k_z=0$. At $\kvec=0$ this expresses conservation of the total $S^z$; at
general $\kvec$ it follows from the projected identity
$P_\mu(S^z_{\mathrm{u}}+S^z_{\mathrm{d}})P_\mu=0$.

The transverse channels have a different interpretation. The lower-layer
$xx$ channel is topologically trivial at leading order and measures the
ordinary $\mu$-magnon spectrum. By contrast, the upper-layer $xx$ channel
contains the two-spinon continuum, but only after convolution with the
$\nu^x$ one-magnon response. The upper-layer $yy$ channel is still more
composite because it probes the $\epsilon$ sector and the $\nu^x$ magnon
simultaneously. These channels may carry sizable spectral weight, but their
thresholds are not direct anyon thresholds.

Representative spectra across these channels are contrasted in
Fig.~\ref{fig:dsf_channels}. For $\mathcal{S}^{zz}_{\mathrm{d}\mathrm{d}}$,
we plot only the connected fluctuation-dressed contribution
$\mathcal{S}_{mm}*\mathcal{S}^{\mathrm{conn}}_{\delta\nu^z\delta\nu^z}/4$,
since its static part simply duplicates $\nu_0^2\mathcal{S}^{zz}_{\mathrm{u}\mathrm{u}}$.
To illustrate the kinematic convolution features, we adopt the $(1,+)$
spinon saddle at $\bar V=10$ ($J^z_{1,n}=2$, $J^z_{2,n}=1$, giving
$\lambda\approx 3.44$ and $\mathcal{B}\approx 0.40$), alongside the
phenomenological vison and $\mu$-magnon parameters from
Fig.~\ref{fig:elementary_dispersions}(c,d). Because the odd vison
background carries $\pi$ flux per dual hexagon, the vison modes are
computed in the magnetic Brillouin zone and folded back into the kagom\'e
Brillouin zone on a $240\times 240$ grid (with Gaussian broadening
$\sigma_\omega=0.08$). This zone folding directly shapes the momentum
profile of the bare two-vison continuum in
$\mathcal{S}^{zz}_{\mathrm{u}\mathrm{u}}$, providing a sharp spectral
benchmark against the broadened, magnon-dressed continua in the composite
channels.

\section{Discussion}
\label{sec7}

The bilayer kagom\'e construction gives a concrete example in which a
topological order is present but is not generically exposed
with simple spin variables nor a simple
fractionalized continuum in the physical spin response. In a conventional
single-variable spin liquid, the spin operator often has a direct overlap with
a pair of fractionalized quasiparticles. Here the topological sector is the
hidden $\tau$ variable, while the measured microscopic spin contains both
$\tau$ and the auxiliary interlayer-alignment variable $\mu$. As a result,
most spin structure factors probe composite operators. A broad continuum in
such a channel need not be a bare spinon or vison continuum, and the absence of a
clean continuum in a generic channel does not by itself exclude a
$\mathbb{Z}_2$ spin liquid.

The most direct layer-resolved signature in this model is the upper-layer
longitudinal response. The operator identity
$S^z_{\mathrm{u}i}=\mathcal{O}^m_i/2$ makes
$\mathcal{S}^{zz}_{\mathrm{u}\mathrm{u}}(\kvec,\omega)$ a direct probe of
the two-vison continuum. The lower-layer longitudinal response contains the
same bare continuum through the static value of $\nu^z$, together with an
additional connected dressing from $\delta\nu^z$ fluctuations. Other channels
contain useful information, but their interpretation requires the auxiliary
$\mu$ spectrum. For example,
$\mathcal{S}^{xx}_{\mathrm{u}\mathrm{u}}$ contains the two-spinon continuum
only after convolution with a one-$\mu$-magnon response, whereas
$\mathcal{S}^{xx}_{\mathrm{d}\mathrm{d}}$ is dominated by the ordinary
$\mu$-magnon branch.

Experimentally, this structure favors probes that can resolve the spin
polarization and layer content. Polarized inelastic neutron scattering is a
natural starting point because it can separate the longitudinal and transverse
spin components. In a bilayer material, the momentum dependence along the
interlayer direction separates the layer combinations directly. By
Eq.~\eqref{eq:bilayer_kz_response}, the low-energy two-vison continuum is
modulated as $\sin^2\!\qty(k_z d/2)$ and vanishes at $k_z=0$. Other
layer-sensitive magnetic spectroscopies could play a similar role if they
couple differently to the two kagom\'e layers. The central experimental
target is not simply a broad continuum, but the contrast between the bare
two-vison continuum in the longitudinal layer-odd channel and the dressed
continua predicted in the remaining channels.

The concrete construction requires the kagom\'e bilayers, a strong local interlayer
Ising coupling, and frustrated longer-range Ising interactions that can
approximate the BFG condition in Eq.~\eqref{eq:bfg_condition}. Candidate
platforms include bilayer kagom\'e transition-metal and rare-earth magnets. A
quantitative material proposal
would require the microscopic exchange estimates, crystal-field analysis, and a
calculation of detailed neutron form factors. Those aspects are outside the scope of the
present theory. The result here is instead a diagnostic principle. Hidden
variables in a multi-component magnet~\cite{Chen_2024} can relocate the topological
continuum to a more convoluted and selected response channel.

In summary, we have constructed a bilayer kagom\'e lattice antiferromagnet in which the
low-energy pseudospin $\tau$ realizes the BFG route to a
$\mathbb{Z}_2$ topological order, while the auxiliary interlayer-alignment variable
$\mu$ remains topologically trivial. The $\tau$-$\mu$ mapping gives an exact
local dictionary between physical spin operators and emergent operators of the
hidden topological theory. This dictionary leads to the concrete spectroscopic
selection rules. $\mathcal{S}^{zz}_{\mathrm{u}\mathrm{u}}$ detects the bare
two-vison continuum, whereas most other diagonal layer-resolved channels
measure convolutions of topological continua and $\mu$-magnon responses.
Rather than advocating the specific and concrete construction in this work,
we actually strongly hope this work to inspire the search and understanding
of exotic and topological phases
among the multi-component frustrated magnets and quantum materials
where the topological orders and features are manifested with the
most visible degrees of freedom, and are instead buried in
the emergent and hidden variables.
The current work is merely an example to illustrate this thinking,
not a detailed and concrete case to be implemented.

\section*{Acknowledgments}

This work was supported by the NSFC (Grant No.~92565110 and 12574061),
the Quantum Science and Technology--National Science and Technology Major Project
(Grant No.~2025ZD0300500), and the BJNSF (Grant No.~F261004).

\bibliography{references}

@article{balents_fractionalization_2002,
  title   = {Fractionalization in an Easy-Axis {{Kagome}} Antiferromagnet},
  author  = {Balents, L. and Fisher, M. P. A. and Girvin, S. M.},
  year    = 2002,
  month   = may,
  journal = {Phys. Rev. B},
  volume  = {65},
  number  = {22},
  pages   = {224412},
  doi     = {10.1103/PhysRevB.65.224412}
}

@article{balents_SpinLiquidsFrustrated_2010,
  title   = {Spin Liquids in Frustrated Magnets},
  author  = {Balents, Leon},
  year    = 2010,
  month   = mar,
  journal = {Nature},
  volume  = {464},
  number  = {7286},
  pages   = {199--208},
  doi     = {10.1038/nature08917}
}

@article{banerjee_NeutronScatteringProximate_2017,
  title   = {Neutron scattering in the proximate quantum spin liquid {$\alpha$-RuCl$_3$}},
  author  = {Banerjee, Arnab and Yan, Jiaqiang and Knolle, Johannes and Bridges, Craig A. and Stone, Matthew B. and Lumsden, Mark D. and Mandrus, David G. and Tennant, David A. and Moessner, Roderich and Nagler, Stephen E.},
  year    = 2017,
  month   = jun,
  journal = {Science},
  volume  = {356},
  number  = {6342},
  pages   = {1055--1059},
  doi     = {10.1126/science.aah6015}
}

@article{becker_DiagnosingFractionalizationSpin_2018,
  title   = {Diagnosing Fractionalization from the Spin Dynamics of {$Z_2$} Spin Liquids on the Kagome Lattice by Quantum {Monte Carlo} Simulations},
  author  = {Becker, Jonas and Wessel, Stefan},
  year    = 2018,
  month   = aug,
  journal = {Phys. Rev. Lett.},
  volume  = {121},
  number  = {7},
  pages   = {077202},
  doi     = {10.1103/PhysRevLett.121.077202}
}

@article{broholm_QuantumSpinLiquids_2020,
  title   = {Quantum Spin Liquids},
  author  = {Broholm, C. and Cava, R. J. and Kivelson, S. A. and Nocera, D. G. and Norman, M. R. and Senthil, T.},
  year    = 2020,
  month   = jan,
  journal = {Science},
  volume  = {367},
  number  = {6475},
  pages   = {eaay0668},
  doi     = {10.1126/science.aay0668}
}

@article{Chen_2024,
  title     = {{Multiflavor Mott insulators in quantum materials and ultracold atoms}},
  volume    = {9},
  issn      = {2397-4648},
  url       = {http://dx.doi.org/10.1038/s41535-023-00614-2},
  doi       = {10.1038/s41535-023-00614-2},
  number    = {1},
  journal   = {npj Quantum Materials},
  publisher = {Springer Science and Business Media LLC},
  author    = {Chen, Gang v. and Wu, Congjun},
  year      = {2024},
  month     = Jan
}

@article{cai_SignaturesFractionalQuantum_2023,
  title   = {Signatures of fractional quantum anomalous {Hall} states in twisted {MoTe$_2$}},
  author  = {Cai, Jiaqi and Anderson, Eric and Wang, Chong and Zhang, Xiaowei and Liu, Xiaoyu and Holtzmann, William and Zhang, Yinong and Fan, Fengren and Taniguchi, Takashi and Watanabe, Kenji and Ran, Ying and Cao, Ting and Fu, Liang and Xiao, Di and Yao, Wang and Xu, Xiaodong},
  year    = 2023,
  month   = oct,
  journal = {Nature},
  volume  = {622},
  number  = {7981},
  pages   = {63--68},
  doi     = {10.1038/s41586-023-06289-w}
}

@article{chen_emergent_2024,
  title   = {Emergent {{Berezinskii-Kosterlitz-Thouless}} and {{Kugel-Khomskii Physics}} in the {{Triangular Lattice Bilayer Colbaltate}}},
  author  = {Chen, G.},
  year    = 2024,
  month   = sep,
  journal = {Phys. Rev. Lett.},
  volume  = {133},
  number  = {13},
  pages   = {136703},
  doi     = {10.1103/PhysRevLett.133.136703}
}

@article{PhysRevResearch.1.033141,
  title     = {{Intrinsic transverse field in frustrated quantum Ising magnets: Physical origin and quantum effects}},
  author    = {Chen, Gang},
  journal   = {Phys. Rev. Res.},
  volume    = {1},
  issue     = {3},
  pages     = {033141},
  numpages  = {9},
  year      = {2019},
  month     = {Dec},
  publisher = {American Physical Society},
  doi       = {10.1103/PhysRevResearch.1.033141},
  url       = {https://link.aps.org/doi/10.1103/PhysRevResearch.1.033141}
}

@article{PhysRevResearch.2.043013,
  title     = {{Intrinsic quantum Ising model on a triangular lattice magnet $\mathrm{Tm}\mathrm{Mg}\mathrm{Ga}{\mathrm{O}}_{4}$}},
  author    = {Liu, Changle and Huang, Chun-Jiong and Chen, Gang},
  journal   = {Phys. Rev. Res.},
  volume    = {2},
  issue     = {4},
  pages     = {043013},
  numpages  = {15},
  year      = {2020},
  month     = {Oct},
  publisher = {American Physical Society},
  doi       = {10.1103/PhysRevResearch.2.043013},
  url       = {https://link.aps.org/doi/10.1103/PhysRevResearch.2.043013}
}

@article{han_FractionalizedExcitationsSpinliquid_2012,
  title   = {Fractionalized Excitations in the Spin-Liquid State of a Kagome-Lattice Antiferromagnet},
  author  = {Han, Tian-Heng and Helton, Joel S. and Chu, Shaoyan and Nocera, Daniel G. and {Rodriguez-Rivera}, Jose A. and Broholm, Collin and Lee, Young S.},
  year    = 2012,
  month   = dec,
  journal = {Nature},
  volume  = {492},
  number  = {7429},
  pages   = {406--410},
  doi     = {10.1038/nature11659}
}

@article{hermele_PyrochlorePhotonsU1_2004,
  title   = {Pyrochlore Photons: {{The U}}(1) Spin Liquid in a {{S}}=1/2 Three-Dimensional Frustrated Magnet},
  author  = {Hermele, Michael and Fisher, Matthew P. A. and Balents, Leon},
  year    = 2004,
  month   = feb,
  journal = {Phys. Rev. B},
  volume  = {69},
  number  = {6},
  pages   = {064404},
  doi     = {10.1103/PhysRevB.69.064404}
}

@article{huh_VisonStatesConfinement_2011,
  title   = {Vison states and confinement transitions of {$\mathbb{Z}_2$} spin liquids on the kagome lattice},
  author  = {Huh, Yejin and Punk, Matthias and Sachdev, Subir},
  year    = 2011,
  month   = sep,
  journal = {Phys. Rev. B},
  volume  = {84},
  number  = {9},
  pages   = {094419},
  doi     = {10.1103/PhysRevB.84.094419}
}

@article{isakov_SpinLiquidPhaseSpin1_2006,
  title   = {Spin-{{Liquid Phase}} in a {{Spin-1}}/2 {{Quantum Magnet}} on the {{Kagome Lattice}}},
  author  = {Isakov, S. V. and Kim, Yong Baek and Paramekanti, A.},
  year    = 2006,
  month   = nov,
  journal = {Phys. Rev. Lett.},
  volume  = {97},
  number  = {20},
  pages   = {207204},
  doi     = {10.1103/PhysRevLett.97.207204}
}

@article{jackeli_MottInsulatorsStrong_2009,
  title   = {Mott {{Insulators}} in the {{Strong Spin-Orbit Coupling Limit}}: {{From Heisenberg}} to a {{Quantum Compass}} and {{Kitaev Models}}},
  author  = {Jackeli, G. and Khaliullin, G.},
  year    = 2009,
  month   = jan,
  journal = {Phys. Rev. Lett.},
  volume  = {102},
  number  = {1},
  pages   = {017205},
  doi     = {10.1103/PhysRevLett.102.017205}
}

@article{kitaev_AnyonsExactlySolved_2006,
  title   = {Anyons in an Exactly Solved Model and Beyond},
  author  = {Kitaev, Alexei},
  year    = 2006,
  month   = jan,
  journal = {Ann. Phys.},
  volume  = {321},
  number  = {1},
  pages   = {2--111},
  doi     = {10.1016/j.aop.2005.10.005}
}

@article{kitaev_FaulttolerantQuantumComputation_2003,
  title   = {Fault-Tolerant Quantum Computation by Anyons},
  author  = {Kitaev, A. {\relax Yu}.},
  year    = 2003,
  month   = jan,
  journal = {Ann. Phys.},
  volume  = {303},
  number  = {1},
  pages   = {2--30},
  doi     = {10.1016/S0003-4916(02)00018-0}
}

@article{knolle_DynamicsTwoDimensionalQuantum_2014,
  title   = {Dynamics of a {{Two-Dimensional Quantum Spin Liquid}}: {{Signatures}} of {{Emergent Majorana Fermions}} and {{Fluxes}}},
  author  = {Knolle, J. and Kovrizhin, D. L. and Chalker, J. T. and Moessner, R.},
  year    = 2014,
  month   = may,
  journal = {Phys. Rev. Lett.},
  volume  = {112},
  number  = {20},
  pages   = {207203},
  doi     = {10.1103/PhysRevLett.112.207203}
}

@article{laughlin_AnomalousQuantumHall_1983,
  title   = {Anomalous {{Quantum Hall Effect}}: {{An Incompressible Quantum Fluid}} with {{Fractionally Charged Excitations}}},
  author  = {Laughlin, R. B.},
  year    = 1983,
  month   = may,
  journal = {Phys. Rev. Lett.},
  volume  = {50},
  number  = {18},
  pages   = {1395--1398},
  doi     = {10.1103/PhysRevLett.50.1395}
}

@article{levin_StringnetCondensationPhysical_2005,
  title   = {String-Net Condensation: {{A}} Physical Mechanism for Topological Phases},
  author  = {Levin, Michael A. and Wen, Xiao-Gang},
  year    = 2005,
  month   = jan,
  journal = {Phys. Rev. B},
  volume  = {71},
  number  = {4},
  pages   = {045110},
  doi     = {10.1103/PhysRevB.71.045110}
}

@article{park_ObservationFractionallyQuantized_2023,
  title   = {Observation of Fractionally Quantized Anomalous {{Hall}} Effect},
  author  = {Park, Heonjoon and Cai, Jiaqi and Anderson, Eric and Zhang, Yinong and Zhu, Jiayi and Liu, Xiaoyu and Wang, Chong and Holtzmann, William and Hu, Chaowei and Liu, Zhaoyu and Taniguchi, Takashi and Watanabe, Kenji and Chu, Jiun-Haw and Cao, Ting and Fu, Liang and Yao, Wang and Chang, Cui-Zu and Cobden, David and Xiao, Di and Xu, Xiaodong},
  year    = 2023,
  month   = oct,
  journal = {Nature},
  volume  = {622},
  number  = {7981},
  pages   = {74--79},
  doi     = {10.1038/s41586-023-06536-0}
}

@article{savary_QuantumSpinLiquids_2017,
  title   = {Quantum Spin Liquids: A Review},
  author  = {Savary, Lucile and Balents, Leon},
  year    = 2017,
  month   = jan,
  journal = {Rep. Prog. Phys.},
  volume  = {80},
  number  = {1},
  pages   = {016502},
  doi     = {10.1088/0034-4885/80/1/016502}
}

@article{tsui_TwoDimensionalMagnetotransportExtreme_1982,
  title   = {Two-{{Dimensional Magnetotransport}} in the {{Extreme Quantum Limit}}},
  author  = {Tsui, D. C. and Stormer, H. L. and Gossard, A. C.},
  year    = 1982,
  month   = may,
  journal = {Phys. Rev. Lett.},
  volume  = {48},
  number  = {22},
  pages   = {1559--1562},
  doi     = {10.1103/PhysRevLett.48.1559}
}

@article{wen_ExperimentalIdentificationQuantum_2019,
  title   = {Experimental Identification of Quantum Spin Liquids},
  author  = {Wen, Jinsheng and Yu, Shun-Li and Li, Shiyan and Yu, Weiqiang and Li, Jian-Xin},
  year    = 2019,
  month   = apr,
  journal = {Npj Quantum Mater.},
  volume  = {4},
  number  = {1},
  pages   = {12},
  doi     = {10.1038/s41535-019-0151-6}
}

@article{wen_TopologicalOrdersRigid_1990,
  title   = {Topological Orders in Rigid States},
  author  = {Wen, Xiao-Gang},
  year    = 1990,
  month   = feb,
  journal = {Int. J. Mod. Phys. B},
  volume  = {4},
  number  = {02},
  pages   = {239--271},
  doi     = {10.1142/S0217979290000139}
}

@article{zeng_ThermodynamicEvidenceFractional_2023,
  title   = {Thermodynamic evidence of fractional {Chern} insulator in moir\'e {MoTe$_2$}},
  author  = {Zeng, Yihang and Xia, Zhengchao and Kang, Kaifei and Zhu, Jiacheng and Kn{\"u}ppel, Patrick and Vaswani, Chirag and Watanabe, Kenji and Taniguchi, Takashi and Mak, Kin Fai and Shan, Jie},
  year    = 2023,
  month   = oct,
  journal = {Nature},
  volume  = {622},
  number  = {7981},
  pages   = {69--73},
  doi     = {10.1038/s41586-023-06452-3}
}

@article{zhao_z_2_2025,
  title   = {{$Z_2$} topological orders in kagome dipolar systems: Feedback from {Rydberg} quantum simulator},
  author  = {Zhao, Pengwei and v. Chen, Gang},
  year    = 2025,
  month   = aug,
  journal = {SciPost Phys.},
  volume  = {19},
  number  = {2},
  pages   = {040},
  doi     = {10.21468/SciPostPhys.19.2.040}
}

\newpage

\appendix

\section{Spinon Mean-Field Theory}
\label{app:spinon_mf}

\begin{table}[b]
  \caption{Onsets of gapped spinon minima in the softened mean-field ansatz.
    The endpoints themselves have zero gap. These are existence boundaries
  for the individual branches, not energy crossings with condensed states.}
  \label{tab:spinon_gapped_onsets}
  \begin{ruledtabular}
    \begin{tabular}{cccc}
      $Q$ & $\operatorname{sign}\mathcal B$ & $\bar V_c/J_0^\perp$ & Soft mode \\
      \hline
      1 & $+$ & 5.939 & $K$ \\
      1 & $-$ & 13.457 & $\Gamma$ \\
      2 & $+$ & 17.315 & $K$ \\
      2 & $-$ & 38.345 & $\Gamma$ \\
    \end{tabular}
  \end{ruledtabular}
\end{table}

Let $n^B_\ell=B_\ell^\dagger B_\ell$ and
$n^b_{\rvec}=b_{\rvec}^\dagger b_{\rvec}$. For a background sign $s=\pm1$,
choose $\kappa^x_{\rvec}=s(1-2n^b_{\rvec})$. The exact Gauss law becomes
\begin{equation}
  (-1)^{n^b_{\rvec}+\sum_{\ell\ni\rvec}n^B_\ell}=s.
\end{equation}
The odd BFG vacuum has $s=-1$, $n^b_{\rvec}=0$, and three occupied $B$
links at every vertex. Here $n^B_i=1/2-\tau_i^z$ counts the complement of
the occupied-link convention in Sec.~III; both counts equal three in the
constrained manifold.

In our mean-field treatment, we approximate $\expval{n^B_\ell}\simeq\mathcal B^2$
and relax the exact parity constraint to the average number condition
$\expval{n^b}+6\mathcal B^2=Q$ with $s=(-1)^Q$. While the $Q=1,2$ ans\"atze
do not strictly enforce the local three-dimer constraint of the BFG
manifold, they serve as tractable variational descriptions for the gapped
spinon bands.

For the scan with fixed $\expval{\mu^z}=-1/2$, let $N_\triangle=N/3$ denote the number of
triangular sites and use a normalized Brillouin-zone average
$\langle f\rangle_{\kvec}=N_\triangle^{-1}\sum_{\kvec}f(\kvec)$.
The constant term retained in Eq.~\eqref{eq:spinon_mf_hamiltonian} is
\begin{equation}
  \frac{E_c}{N_\triangle}
  =\frac34\bar V(1-2\mathcal B^2)^2
  -\frac{3J_0^z}{4}+\lambda(6\mathcal B^2-Q).
  \label{eq:spinon_constant_energy}
\end{equation}
Writing $A_{\kvec}=\lambda+J_e\gamma_{\kvec}$ and
$D_{\kvec}=J_e\gamma_{\kvec}$ gives
$E_e=\sqrt{A_{\kvec}^2-D_{\kvec}^2}$ and vacuum occupation
$v_{\kvec}^2=A_{\kvec}/(2E_e)-1/2$. The ground-state energy density is
\begin{equation}
  f=\frac{E_c}{N_\triangle}
  +\frac12\langle E_e(\kvec)-\lambda\rangle_{\kvec},
\end{equation}
where $\langle\gamma_{\kvec}\rangle_{\kvec}=0$ has been used. Stationarity
with respect to $\lambda$ and $\mathcal B$, at $J_e=J_0^\perp\mathcal B$, gives
\begin{equation}
  \begin{aligned}
    0&=\left\langle\frac{\lambda+J_e\gamma_{\kvec}}{2E_e(\kvec)}
    -\frac12\right\rangle_{\kvec}+6\mathcal B^2-Q,\\
    0&=\frac{J_0^\perp\lambda}{2}
    \left\langle\frac{\gamma_{\kvec}}{E_e(\kvec)}\right\rangle_{\kvec}
    +12\lambda\mathcal B
    +6\bar V\mathcal B(2\mathcal B^2-1).
  \end{aligned}
  \label{eq:spinon_saddle_equations}
\end{equation}
We select the lowest $f$ among gapped, uncondensed constrained minima.
The multiplier
$\lambda$ enforces the number constraint and is not an independent
energy-minimization variable. Writing its solution as $\lambda(\mathcal B)$,
we minimize $f_{\mathrm{red}}(\mathcal B)=f(\lambda(\mathcal B),\mathcal B)$
and require positive curvature along this curve.

A convenient parametrization is $x=J_e/\lambda$. Positivity of the bosonic
quadratic Hamiltonian requires $\lambda>0$ and $-1/12<x<1/6$. Define the
thermodynamic-limit Brillouin-zone averages
\begin{equation}
  \begin{aligned}
    I_\pm(x)&=\left\langle(1+2x\gamma_{\kvec})^{\pm1/2}
    \right\rangle_{\kvec},\\
    n(x)&=\frac{I_+(x)+I_-(x)}{4}-\frac12.
  \end{aligned}
  \label{eq:spinon_continuum_moments}
\end{equation}
The number constraint then determines
\begin{equation}
  \mathcal B=\operatorname{sign}(x)\sqrt{\frac{Q-n(x)}6},\qquad
  \lambda=\frac{J_0^\perp\mathcal B}{x},
\end{equation}
and the energy on the constraint curve becomes
\begin{equation}
  f_{\mathrm{red}}
  =\frac34\bar V(1-2\mathcal B^2)^2-\frac{3J_0^z}{4}
  +\frac\lambda4\qty[I_+(x)-I_-(x)].
\end{equation}
One reciprocal coordinate can be integrated analytically in terms of
complete elliptic integrals, leaving a one-dimensional quadrature. Near the
soft-mode boundary $x_c=1/6$ or $-1/12$, we use
$\log(1-x/x_c)$ to resolve the small positive gap. Direct minimization of
$f_{\mathrm{red}}$ is followed by refinement with
Eq.~\eqref{eq:spinon_saddle_equations} and a check of the constrained
curvature.

At $x=x_c$, the regular-mode density $n_c=n(x_c)$ is finite. Let
$b_c=\sqrt{(Q-n_c)/6}$ and $c=6$ for $\mathcal B>0$ or $12$ for
$\mathcal B<0$. The gapped branch starts at
\begin{equation}
  \frac{\bar V_c}{J_0^\perp}
  =\frac{c\qty[18b_c^2-Q+\frac12\qty(I_+(x_c)-1)]}
  {6b_c(1-2b_c^2)}.
  \label{eq:spinon_gapped_onset}
\end{equation}
The resulting values are given in Table~\ref{tab:spinon_gapped_onsets}.
For $\bar V>\bar V_c$, the corresponding branch has a gapped interior
minimum. At the onset its gap closes at $K$ for $\mathcal B>0$ or at
$\Gamma$ for $\mathcal B<0$. Continuing below the onset requires treating
the soft-mode occupation separately; such condensed branches are outside
the gapped variational space considered here.

\section{Vison Duality and Soft-Spin Approximation}
\label{app:vison_duality}

\subsection{Exact duality}
\label{app:duality_exact}

Let $\ell=\nnb{\rvec\rvec'}$ be a triangular-lattice link and
$\ell^*=\nnb{\vb*{R}\vb*{R}'}$ its crossing honeycomb link.
The direct triangle $\triangle_{\vb*{R}}$ is centered at $\vb*{R}$,
and the dual hexagon $\hexagon^*_{\rvec}$ surrounds the direct vertex $\rvec$.
Here $\partial$ denotes the boundary of a plaquette. Products over
$\partial\triangle_{\vb*{R}}$ and $\partial\hexagon^*_{\rvec}$ run over
the three and six links forming the respective boundaries.
All four sets of variables $\sigma,\kappa,\eta,\pi$ are Pauli operators.
The original physical Hilbert space obeys
\begin{equation}
  G_{\rvec}\equiv\kappa^x_{\rvec}
  \prod_{\ell\ni\rvec}\sigma^x_\ell=1.
  \label{eq:direct_gauss_generator}
\end{equation}
Within the corresponding physical subspaces, we introduce the duality mapping
\begin{equation}
  \begin{aligned}
    \eta^x_{\ell^*}
    &=\kappa^z_{\rvec}\sigma^z_\ell\kappa^z_{\rvec'},\quad
    \sigma^x_\ell
    &=\pi^z_{\vb*{R}}\eta^z_{\ell^*}\pi^z_{\vb*{R}'},\\
    \pi^x_{\vb*{R}}
    &=\prod_{\ell\in\partial\triangle_{\vb*{R}}}\sigma^z_\ell,\quad
    \kappa^x_{\rvec}
    &=\prod_{\ell^*\in\partial\hexagon^*_{\rvec}}\eta^z_{\ell^*}.
  \end{aligned}
  \label{eq:full_gauge_duality}
\end{equation}
The first two lines exchange the gauge-invariant Pauli pair on each
crossing link. Both operators square to one, anticommute on that link,
and commute on distinct links. They also commute with their respective
Gauss generators. This mapping therefore preserves the exact gauge-invariant operator algebra
within the physical subspace.

The plaquette and star relations are consistent with the Gauss laws.
Around a direct triangle, each endpoint factor $\kappa^z$ appears twice
and cancels, so the first line gives
\begin{equation}
  \prod_{\ell^*\ni\vb*{R}}\eta^x_{\ell^*}
  =\prod_{\ell\in\partial\triangle_{\vb*{R}}}\sigma^z_\ell
  =\pi^x_{\vb*{R}}.
\end{equation}
This fixes the dual physical subspace through the Gauss law
\begin{equation}
  \widetilde G_{\vb*{R}}\equiv\pi^x_{\vb*{R}}
  \prod_{\ell^*\ni\vb*{R}}\eta^x_{\ell^*}=1,
  \label{eq:dual_gauss_generator}
\end{equation}
where the product contains the three honeycomb links incident on
$\vb*{R}$. Conversely, multiplying the second line of
Eq.~\eqref{eq:full_gauge_duality} over the six links touching $\rvec$
cancels every $\pi^z$ factor on the surrounding dual hexagon. Equation
\eqref{eq:direct_gauss_generator} then gives the fourth line.
Original electric charge is therefore dual magnetic flux, while original
magnetic flux is dual electric charge. Excitations are defined as changes
relative to the chosen charge and flux backgrounds.

The dual redundancy acts as
\begin{equation}
  \begin{aligned}
    \pi^z_{\vb*{R}}&\longmapsto s_{\vb*{R}}\pi^z_{\vb*{R}},\\
    \eta^z_{\vb*{R}\vb*{R}'}&\longmapsto
    s_{\vb*{R}}\eta^z_{\vb*{R}\vb*{R}'}s_{\vb*{R}'},
  \end{aligned}
\end{equation}
with $s_{\vb*{R}}=\pm1$. A single $\pi^z$ is gauge dependent, whereas the
dressed pair in Eq.~\eqref{eq:vison_duality} is gauge invariant. More
generally, an open product of $\sigma^x$ along a dual path becomes a vison
pair joined by an $\eta^z$ Wilson line. An open spinon Wilson line with
$\kappa^z$ endpoints becomes the corresponding product of $\eta^x$.
The two string operators anticommute when their paths cross an odd number
of times, expressing the mutual $\pi$ statistical phase of spinons and visons.

The local identities do not fix the noncontractible loop operators on a
torus. Their algebra must also be carried through the mapping, and any
restriction to a fixed background must keep track of its global holonomies.
The closed-lattice identities
$\prod_{\rvec}\kappa^x_{\rvec}=1$ and
$\prod_{\vb*{R}}\pi^x_{\vb*{R}}=1$ enforce even total charge and flux parity.
Keeping track of these global topological sectors ensures an exact
one-to-one correspondence between the direct and dual Hilbert spaces.

\subsection{Dual Hamiltonian and charge backgrounds}

The duality acts on the $\tau$ sector; all auxiliary $\mu$ operators are
unchanged. The full $\tau$--$\mu$ Hamiltonian can be transformed using the
same operator dictionary. Here we display its leading projected form.
For the kagom\'e lattice site $i$ corresponding to $\ell_i^*$, abbreviate the dressed
dual bond by $Y_i=\pi^z_{\vb*{R}}\eta^z_{\ell_i^*}\pi^z_{\vb*{R}'}$.
Substitution into Eq.~\eqref{eq:effective_tau_tfim} gives the exact dual
form of the projected Hamiltonian,
\begin{equation}
  H_{\mathrm{dual}}
  =\frac14\sum_{i<j}V_{ij}Y_iY_j
  +\frac{J_0^\perp}{2}\sum_i\eta^x_{\ell_i^*}
  -\frac{NJ_0^z}{4},
  \label{eq:projected_dual_hamiltonian}
\end{equation}
subject to Eq.~\eqref{eq:dual_gauss_generator}, where each kagom\'e pair is counted once. Equation~\eqref{eq:projected_dual_hamiltonian}
retains the full dual gauge dynamics, providing the exact starting point
before introducing the static-flux and soft-spin approximations.

To describe visons in a static charge background, we fix a gauge and
replace the original charges by $\kappa^x_{\rvec}\to q_{\rvec}$.
The background links must then satisfy
\begin{equation}
  \prod_{\ell^*\in\partial\hexagon^*_{\rvec}}\bar\eta^z_{\ell^*}
  =q_{\rvec}.
  \label{eq:dual_background_charge_flux}
\end{equation}
The three-dimer BFG constraint gives $q_{\rvec}=-1$, producing one
$\pi$ flux per dual hexagon. This establishes
Eq.~\eqref{eq:odd_vison_flux} directly
from the original Gauss law. On a closed periodic lattice the chosen
charges must obey $\prod_{\rvec}q_{\rvec}=1$; the loop sectors remain
additional data when selecting a background representative.

For an unordered kagom\'e pair, the Ising interaction is
$V_{ij}\tau_i^z\tau_j^z=(V_{ij}/4)\sigma_i^x\sigma_j^x$.
If the corresponding dual links share a site, Eq.~\eqref{eq:vison_duality}
therefore yields
\begin{equation}
  V_{ij}\tau_i^z\tau_j^z
  =\frac{V_{ij}}{4}\eta^z_{\vb*{R}\vb*{R}''}
  \eta^z_{\vb*{R}''\vb*{R}'}\pi^z_{\vb*{R}}\pi^z_{\vb*{R}'},
\end{equation}
using $(\pi^z_{\vb*{R}''})^2=1$ before making a soft-spin approximation.
Thus the bare contribution from a kagom\'e nearest-neighbor bond is
$J_2^\pi=V_1/4$ in the unordered-pair convention of
Eq.~\eqref{eq:vison_soft_hamiltonian_main}. The second- and intrahexagon
third-neighbor terms involve disjoint dual links and remain quartic.

Because all three shells have equal strength at the BFG point, these
quartic terms are comparable to $J_2^\pi$, motivating a phenomenological
treatment of the effective quadratic vison hopping. Furthermore, in the
exact Pauli representation, the nearest-neighbor term
$J_1^\pi\sum_{\ell^*}\eta^z_{\ell^*}\pi^z_{\vb*{R}}\pi^z_{\vb*{R}'}
=2J_1^\pi\sum_i\tau_i^z$ corresponds to a longitudinal field that is odd
under layer exchange. We nevertheless include a phenomenological $J_1^\pi$
in Fig.~\ref{fig:elementary_dispersions}(c) to illustrate the generic
dispersion of the four-band odd-flux vison kernel.

\subsection{Soft-spin fields and quadratic kernels}

Up to this point, $\pi^z$ is a Pauli operator with eigenvalues $\pm1$ and
the local identity $(\pi^z)^2=1$. The soft-spin approximation replaces it
by a real field $\phi_{\vb*{R}}\in\mathbb R$, transforming as
$\phi_{\vb*{R}}\mapsto s_{\vb*{R}}\phi_{\vb*{R}}$ under the same dual
gauge redundancy. Its amplitude is allowed to fluctuate, so
$\phi_{\vb*{R}}^2=1$ is no longer an operator identity.
This replacement retains fluctuations of the vison field and should not
be read as taking the expectation value of $\pi^z$.

Consequently, the shared-site cancellation used above must be carried out
in the Pauli algebra before softening. Replacing all factors by $\phi$
first would leave an intermediate $\phi_{\vb*{R}''}^2$ and would not give
the same quadratic interaction. A spherical approximation can impose an
average length constraint through $\lambda_\pi$ once the dynamics is
specified; it does not restore the exact local two-level Hilbert space.

Writing the static quadratic energy as
$\sum_{\kvec}\phi_{-\kvec}^{T}K(\kvec)\phi_{\kvec}$, an odd-background gauge
with four sites per magnetic unit cell has
\begin{equation}
  K_{\mathrm{odd}}=\lambda_\pi I_4+\frac{J_1^\pi}{2}M_1
  +\frac{J_2^\pi}{2}(M_1^2-3I_4),
\end{equation}
where, with the magnetic Bravais vectors $\vb*{a}_{1,2}^{m}$ and
$z=\ee^{\ii\kvec\cdot\vb*{a}_1^m}$,
$w=\ee^{\ii\kvec\cdot\vb*{a}_2^m}$,
\begin{equation}
  M_1=
  \begin{pmatrix}
    0&z-1&z&0\\
    z^*-1&0&0&w/z\\
    z^*&0&0&1+z^*\\
    0&z/w&1+z&0
  \end{pmatrix}.
\end{equation}
The matrix $M_1^2-3I_4$ sums two-link paths with their gauge signs.
Let $\vb*{a}_{1,2}$ be the triangular Bravais vectors.
We choose $\vb*{a}_1^m=\vb*{a}_2$ and
$\vb*{a}_2^m=2(\vb*{a}_2-\vb*{a}_1)$, so the magnetic unit cell has twice
the area of the original honeycomb unit cell.
The four eigenvalues of $K_{\mathrm{odd}}$ give the static vison bands.

To turn such a kernel into a quantum soft-spin theory, one possible
normalization of the Euclidean action is
\begin{equation}
  S_m=\int\dd\tau\sum_{\kvec}\qty[
    \chi_\pi\,\partial_\tau\phi_{-\kvec}^{T}
    \partial_\tau\phi_{\kvec}
  +\phi_{-\kvec}^{T}K(\kvec)\phi_{\kvec}],
\end{equation}
which yields $\omega_n^m=\sqrt{\varepsilon_n^m/\chi_\pi}$ for
$\chi_\pi>0$ and a positive kernel. For the spectral parametrization used
in Sec.~VI, we adopt $\omega_n^m=c_\pi\varepsilon_n^m$. This linear
assignment is an independent phenomenological choice. In this
parametrization, $\lambda_\pi$ is also specified as an input rather than
determined by a dynamical spherical constraint.

\section{Auxiliary \texorpdfstring{$\mu$}{mu}-Magnon Spin-Wave Theory}
\label{app:mu_spinwave}

For real, translation-invariant bond correlators, write the quadratic
real-space Hamiltonian as
\begin{equation}
  \begin{aligned}
    H_\mu^{(2)}={}&\epsilon_\mu\sum_i a_i^\dagger a_i\\
    &+\sum_{n=1}^3\sum_{\langle ij\rangle_n}
    \qty[t_n^\mu a_i^\dagger a_j+d_n^\mu a_i a_j+\mathrm{H.c.}]
    +E_\mu^{(0)}.
  \end{aligned}
  \label{eq:mu_realspace_quadratic}
\end{equation}
Each pair is counted once, and the third-neighbor sum includes only
intrahexagon pairs. Within the factorization used in Sec.~IV, let
$C_n^{zz}=\expval{\tau_i^z\tau_j^z}$ and
$C_n^\perp=\expval{\tau_i^x\tau_j^x+\tau_i^y\tau_j^y}$ on shell $n$.
When mixed $y$--$z$ correlators vanish, the microscopic transverse terms give
\begin{equation}
  \begin{aligned}
    t_n^\mu&=J_{1,n}^\perp\qty(\frac14+C_n^\perp+C_n^{zz})
    +J_{2,n}^\perp\expval{\tau^x},\\
    d_n^\mu&=J_{1,n}^\perp\qty(\frac14+C_n^\perp-C_n^{zz})
    +J_{2,n}^\perp\expval{\tau^x} ,
  \end{aligned}
\end{equation}
with the on-site energy $\epsilon_\mu$ given by Eq.~\eqref{eq:mu_gap_formula}.
In Fig.~\ref{fig:elementary_dispersions}(d) and Sec.~\ref{sec6}, we treat
$\epsilon_\mu$, $t_n^\mu$, and $d_n^\mu$ as effective phenomenological
parameters to illustrate the magnon dispersion and its convolution with the
topological continua.

Let $\mathcal N_n(a,b)$ be the set of cell displacements connecting
sublattices $a,b$ by a shell-$n$ bond. Define
\begin{equation}
  [F_n(\kvec)]_{ab}=\sum_{\vb*{R}\in\mathcal N_n(a,b)}
  \ee^{\ii\kvec\cdot(\vb*{R}+\vb*{x}_b-\vb*{x}_a)}.
\end{equation}
The normal and anomalous blocks are
\begin{equation}
  A_\mu=\epsilon_\mu I_3+\sum_n t_n^\mu F_n,\qquad
  D_\mu=\sum_n d_n^\mu F_n.
\end{equation}
Using the Nambu convention in Eq.~\eqref{eq:mu_nambu_spinor}, the matrix is
\begin{equation}
  \mathcal H_\mu(\kvec)=
  \begin{pmatrix}
    A_\mu(\kvec)&D_\mu(\kvec)\\
    D_\mu^\dagger(\kvec)&A_\mu^T(-\kvec)
  \end{pmatrix}.
\end{equation}
For a positive-definite $\mathcal H_\mu$, para-unitary diagonalization gives
the three positive eigenvalues of
$\Sigma_z\mathcal H_\mu$, with $\Sigma_z=\mathrm{diag}(I_3,-I_3)$.
The limit $t_n^\mu=d_n^\mu=0$ gives three flat modes of energy
$\epsilon_\mu$, which fixes the factor of $1/2$ in
Eq.~\eqref{eq:mu_bdg_main}.

\end{document}